\documentclass[format=acmsmall]{acmart}

\usepackage{comment}
\usepackage{tabularx}
\usepackage{graphicx}
\usepackage{url}
\usepackage{hyperref}
\usepackage[absolute,overlay]{textpos}
\usepackage{multirow}

\usepackage{color}

\usepackage{xspace}
\usepackage{xcolor}

\newcommand{\ie}{\mbox{i.e.,}\xspace}
\newcommand{\eg}{\mbox{e.g.,}\xspace}

\newcommand{\PDF}{\mbox{\emph{PoP~1}}\xspace}
\newcommand{\PUL}{\mbox{\emph{PoP~2}}\xspace}
\newcommand{\UMB}{\mbox{\emph{PoP~3}}\xspace}

\usepackage{lib/chronosys/chronosys}

\usepackage{subfig}

\newcommand{\nl}{& \\[0.75ex]}

\usepackage{tikz}
\usetikzlibrary{calc,shapes.symbols,shapes.geometric,positioning,arrows,chains}
\usetikzlibrary{decorations,decorations.text,decorations.pathreplacing}

\usepackage{array}
\newcolumntype{x}[1]{>{\centering\let\newline\\\arraybackslash\hspace{0pt}}m{#1}}
\newcolumntype{y}[1]{>{\let\newline\\\arraybackslash\hspace{0pt}}m{#1}}

\begin{document}

\title[You, the Web and Your Device: Longitudinal Characterization of Browsing Habits]
      {You, the Web and Your Device: \\Longitudinal Characterization of Browsing Habits}

\author{Luca Vassio}
\email{luca.vassio@polito.it}
\affiliation{%
  \institution{Politecnico di Torino}
  \country{Italy}
}

\author{Idilio Drago}
\email{idilio.drago@polito.it}
\affiliation{%
  \institution{Politecnico di Torino}
  \country{Italy}
}

\author{Marco Mellia}
\email{marco.mellia@polito.it}
\affiliation{%
  \institution{Politecnico di Torino}
  \country{Italy}
}

\author{Zied Ben Houidi}
\email{zied.ben\_houidi@nokia-bell-labs.com}
\affiliation{%
  \institution{Nokia Bell Labs}
  \country{France}
}

\author{Mohamed Lamine Lamali}
\email{mohamed-lamine.lamali@u-bordeaux.fr}
\affiliation{%
  \institution{LaBRI, Universit\'{e} de Bordeaux}
  \country{France}
}

\begin{abstract}
Understanding how people interact with the web is key for a variety of applications -- \eg from the design of effective web pages to the definition of successful online marketing campaigns. Browsing behavior has been traditionally represented and studied by means of \emph{clickstreams}, i.e., graphs whose vertices are web pages, and edges are the paths followed by users. Obtaining large and representative data to extract clickstreams is however challenging.

The evolution of the web questions whether browsing behavior is changing and, by consequence, whether properties of clickstreams are changing. This paper presents a longitudinal study of clickstreams in from 2013 to 2016. We evaluate an anonymized dataset of HTTP traces captured in a large ISP, where thousands of households are connected. We first propose a methodology to identify actual URLs requested by users from the massive set of requests automatically fired by browsers when rendering web pages. Then, we characterize web usage patterns and clickstreams, taking into account both the temporal evolution and the impact of the device used to explore the web.
Our analyses precisely quantify various aspects of clickstreams and uncover interesting patterns, such as the typical short paths followed by people while navigating the web, the fast increasing trend in browsing from mobile devices and the different roles of search engines and social networks in promoting content.

Finally, we contribute a dataset of anonymized clickstreams to the community to foster new studies.\footnote{Anonymized clickstreams are available to the public at \url{http://bigdata.polito.it/clickstream}}

\end{abstract}

%
%


\thanks{The research leading to these results has been partly funded by the
  Vienna Science and Technology Fund (WWTF) through project ICT15-129, BigDAMA.}

\maketitle

\TPshowboxestrue
\TPMargin{0.3cm}
\begin{textblock*}{14cm}(1.8cm,0.7cm)
	\bf
	\definecolor{myRed}{rgb}{0.55,0,0}
	\color{myRed}
	\noindent
	Please cite this paper as:
Luca Vassio, Idilio Drago, Marco Mellia, Zied Ben Houidi, and Mohamed Lamine Lamali. 2018. You, the Web, and Your Device: Longitudinal Characterization of Browsing Habits. ACM Trans. Web 12, 4, Article 24 (November 2018), 30 pages. 
	\url{https://doi.org/10.1145/3231466}
\end{textblock*}

\renewcommand{\shortauthors}{L. Vassio et al.}

\section{Introduction}
\label{sec:intro}

Since its introduction, the web has become the preferred means to access online information. Technology has evolved, from simple static web pages to dynamic applications that let users search for content, buy goods, spend time on social networks, i.e., to ``browse'' the web. Understanding how people interact with the web has been always a fascinating problem~\cite{bucklin_click_2009,kammenhuber_web_2006,broder_graph_2000,meusel_graph_2014} for a variety of purposes like improving search engines~\cite{bai_discovering_2011}, comparing rankings in the web~\cite{craswell_random_2007,joachims_optimizing_2002}, recommending content to people~\cite{Mele:2013:WUM:2433396.2433493}, or increasing privacy and security~\cite{wang2013you}.


Browsing activities have been typically modeled using graphs -- or clickstream graphs as they are typically called -- where vertices are the visited web pages, and edges are the followed hyperlinks. They capture the paths that users take when navigating through websites. The evolution of the web, obviously, changes how users interact with it. We today witness the predominance of a handful of popular online services~\cite{gehlen_uncovering_2012} and the rise of mobile devices. How are these factors changing the way we browse the web?

In this paper we are interested in answering the following two questions:
\begin{itemize}
\item How are the clickstream graphs affected by the web evolution over the past years?
\item What are the differences between clickstream graphs from different browsing devices (\eg PCs and smartphones)?
\end{itemize}

We provide a longitudinal characterization of the clickstream graphs. Fundamental to answer these questions is the availability of data. Previous studies are either outdated~\cite{huberman_strong_1998}, or focused on small sets of volunteers~\cite{tossell_characterizing_2012,sellen_how_2002,weinreich_not_2008}, on user interactions with search engines~\cite{DuarteTorres:2014:ASB:2600093.2555595} or proxy logs~\cite{adar_large_2008}. Only few studies have used passive network traces to study browsing behavior~\cite{kammenhuber_web_2006,xie_resurf_2013}. In this paper, we leverage a three-year long anonymized dataset of passive measurements from an ISP network, thus offering a privileged observation point. The probes monitored 25\,000 \emph{households} and observed more than 64 billion HTTP requests. From this dataset, we extract the subset of requests related to explicitly visited web pages (user-actions) from the mass of objects automatically fetched by browsers (automatic-actions). This is a complicated task, since modern web pages are complex and include many HTML files, JavaScript, and multimedia objects. We tackle it by adopting a novel approach based on machine learning algorithms.

Given the requested web pages and the hyperlinks followed by people, we build \emph{clickstream graphs for each browser in a household}. In total, we have 5.5 million graphs corresponding to over 1 billion visited web pages. We then exploit this dataset to investigate browsing habits, providing the evolution over three years, and carefully characterizing differences in usage according to device types. In summary, this paper makes three contributions:

\begin{itemize}
\item We propose a new approach for the identification of web pages explicitly visited by users in HTTP logs collected by passive network monitors. Our approach generalizes ad-hoc designed heuristics~\cite{ihm_towards_2011,xie_resurf_2013,houidi_gold_2014}, automatically learning the patterns that characterize explicit visits, with detection precision and recall over 90\%.

\item We present a characterization of clickstreams that differs from previous efforts~\cite{xie_resurf_2013,kumar_characterization_2010} for (i)~covering a large population during a long consecutive period of time, and (ii)~accounting for different device types used to browser the web at home.

\item We contribute to the community a three-year long dataset of anonymized clickstreams, covering thousands of households in Europe. To the best of our knowledge, this is one of the largest datasets that includes clickstream graphs from regular internet users, browsing with multiple devices.

\end{itemize}

We focus on global patterns, highlighting when there are general trends in surfing habits, rather than unexpected but rare events. Our analysis confirms and precisely quantifies many intuitions about the way people navigate the web, besides leading to a number of interesting findings:

\begin{itemize}

\item Search Engines (SEs) and Online Social Networks (OSNs) are among the preferred means to discover content. As of 2016, 54\% of web domains were visited starting from Google, and 9\% (6\% in 2013) starting from Facebook. SEs are starting point of longer and deeper navigation, while content promoted by OSNs typically generates visits to a single or very few web pages. Interestingly, OSNs are much more important to discover content on smartphones than on PCs, a result previously not highlighted.

\item Web page complexity has continuously increased from 2013 to 2016, with URLs intentionally visited by users going from 2\% to 1.5\% of the total number of URLs requested by browsers.

\item The number of devices and applications used to browse the web at home has increased significantly, with smartphones and tablets accounting for 29\% and 9\% of the visited web pages in 2016, respectively.  Users are interacting more frequently with the web from their smartphones at home than in the past~\cite{xie_resurf_2013}. However, in a session on a mobile app on average  only $5$ web pages are visited, in a time span of only 2 minutes.

\item When considering the number of visited web pages, we observe that 50\% of the clickstream graphs include less than 27 web pages per day for PCs (8 for smartphones), belonging to less than 9 domains (4 for smartphones). Considering consecutive visited web pages, \ie a path, we observe that people stay in very few domains, even when
navigating through hundreds of web pages. These numbers have mostly remained constant over the years, despite changes in devices and applications used to browse the web.

\item Encryption has gained momentum in the web with many popular domains migrating to HTTPS. We clearly see the impact of HTTPS on properties of the clickstream graphs. Still, in June 2016, only around 13\% of the domains are served (partly or totally) in HTTPS, and 85\% of the encrypted traffic is related to the top-20 content providers, like Google and Facebook. Through data-driven simulations, we also provide estimations for the impact of HTTPS migration on the clickstream graphs.

\end{itemize}

Findings and the contributed dataset have several implications to the Internet actors. For example, they can (i)~help advertisers to make informed decisions on whether to target ads campaigns on mobile or PC users; (ii)~help network operators to understand interests of users and recommend products and services; (iii)~help researchers to investigate privacy aspects related to properties of clickstreams learned from traffic; and, more generally, (iv)~help the research community to study the place of web technologies in people's life.

\subsection{Scope and Limitations}

The scope of our study is obviously limited by the coverage and characteristics of the dataset.

First, the evaluated dataset is limited to the non-encrypted part of the web. It however covers a particularly interesting period, in which the usage of HTTPS has grown from negligible to noticeable percentages. We analyze from different points of view the impact of encryption on our dataset. Despite the growth on HTTPS usage, the majority of the domains were still not encrypted by the end of the data capture in 2016. Moreover, transitions {\it from} popular encrypted domains {\it to} the unencrypted ones are still visible in the analysis. This happens because early adopters of full HTTPS deployments are large content promoters (\eg Google and Facebook) that still inform non-encrypted domains the origin of visits. As such, we have no information about actions performed \emph{inside} these encrypted domains, but we see the transition when users eventually leave them towards unencrypted domains. An encrypted domain appears as a single vertex in a clickstream graph, connected to all vertices representing domains visited from it.

Second, no information to identify people exists in the dataset. Section~\ref{sec:datasets} will show that households (\ie home internet installations) are identified by anonymized keys, and browsers by \texttt{user agent} strings. Analyses are thus performed in a \emph{per-browser} level -- \ie each \texttt{user agent} string observed in a household. Naturally, people use several browsers to explore the web, and several persons are aggregated in a household. Privacy requirements however limit any different granularity.

Third, the evaluated dataset includes only a regional sample of households in Europe. Users in other regions may have diverse browsing habits that result in different clickstreams. Equally, mobile devices have been monitored only while connected to home WiFi networks. As such, our quantification of browsing on mobile terminals is actually a lower-bound, since visits while connected to other technologies are not captured.

Last, as in any large-scale analysis of real-world measurements, many preparation steps have proven essential to clean up spurious data and reduce biases on results. For instance, we have observed non-standard implementations of HTTP protocols by some browsers that prevent the reconstruction of the clickstreams in certain situations. Equally, we were faced with many challenges to reconstruct clickstreams on mobile terminals, given the diverse ways that mobile apps operate. We will elaborate further about these technical aspects in Section~\ref{sec:datasets} and on Appendix~\ref{appendix:a}.

\subsection{Reading Map}

After defining the problem, introducing the terminology used throughout the paper (Section~\ref{sec:probTerm}) and discussing related works (Section~\ref{sec:related}), we describe (Section~\ref{sec:validation}) and assess performance (Section~\ref{sec:clasPerf}) of the classifier used to extract visited web pages from raw HTTP logs.
We then apply it to a longitudinal dataset (described in Section~\ref{sec:datasets}) and characterize the browsing habits (Section~\ref{sec:ContentConsumption}), clickstreams (Section~\ref{sec:clickstream}) and content promoters (Section~\ref{sec:content_discovery}). We then conclude the paper (Section~\ref{sec:future}) and present some final technical details about the extraction of clickstreams from HTTPS to HTTP transitions (Appendix~\ref{appendix:a}).
\section{Problem and Terminology}
\label{sec:probTerm}

The key terminology used throughout the paper is summarized in Table~\ref{tab.nom}.

\begin{table}[!h]
	\centering
	\caption{Summary of the key terminology used throughout the paper.}
	\label{tab.nom}
	{
		\small

		\begin{tabular}{|x{2cm}|y{10cm}|x{0cm}}
      \cline{1-2}
			User-action & The initial HTTP request triggered by a user interaction with a browser. Informally, a user-action is a click or a visit to a web page. \nl

			\cline{1-2}
			Web page    & A URL without parameters (\eg \texttt{http://www.example.com/index.html}). Many user-actions thus can be related to the same web page. \nl

			\cline{1-2}
			Domain      & The pair formed by the second-level domain and the top-level domain of a URL (\eg \texttt{example.com}). Informally, a domains is a website. \nl

			\cline{1-2}
      Clickstream & An ordered list of user-actions from a browser that can be represented through a graph.\nl

			\cline{1-2}
			Browser     & An application, identified by its user agent string, used in a \textit{household} to navigate through multiple web pages and domains. \nl

			\cline{1-2}
		\end{tabular}
	}
\end{table}

We are interested in studying how people surf the web, based on the analysis of raw HTTP logs. One challenge is to extract \emph{user-actions} from such noisy logs. Indeed, rendering a web page is a rather complex process that requires the browser to download HTML files, JavaScript, multimedia objects and dynamically generated content.
All these objects are retrieved by the browser by means of independent HTTP requests. We define the initial HTTP request triggered by a user interaction with the browser as a \emph{user-action}, and as \emph{automatic-actions} all the remaining HTTP requests fired to render web pages.

User-actions thus correspond to \emph{web pages} explicitly visited by a user (\eg \texttt{http://www.example.com/index.html}), and the two terms will be interchangeable in this paper. For each web page, we know its second-level domain (\ie \texttt{example}) and top-level domain (\ie \texttt{com}), which combination (\eg \texttt{example.com}) we will simply call a \emph{domain}.

\begin{figure}
  \centering
  \includegraphics[width=0.7\textwidth]{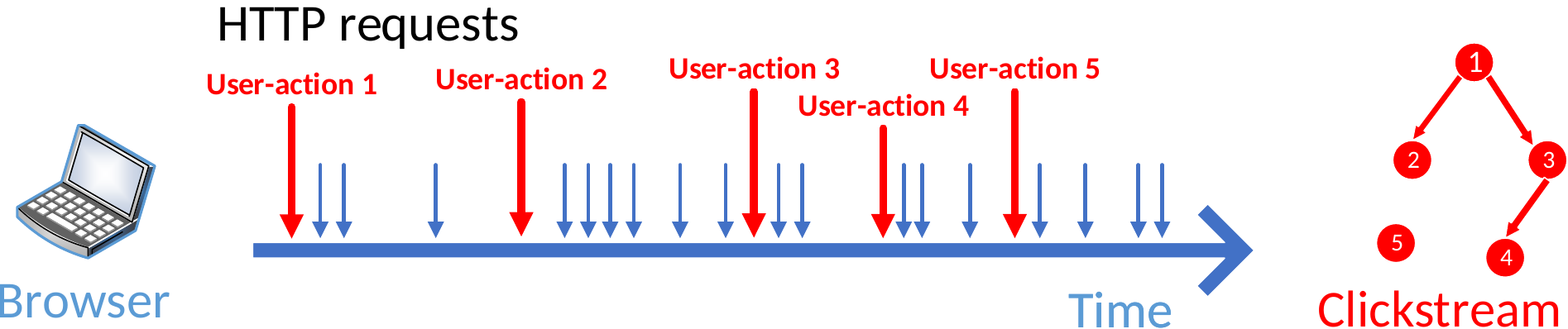}
  \caption{Example of a client browser activity and its relation to our terminology.}
  \label{fig:classificationProblem}
\end{figure}

Figure~\ref{fig:classificationProblem} illustrates these definitions. It depicts the timeline of a user surfing the web. The user visits five web pages, whose corresponding user-actions are marked by tall red arrows. Following each user-action, the browser fires automatic-actions to fetch objects, which we mark with short blue arrows.

We aim at characterizing browsing habits and check how they are evolving. We call the \emph{clickstream} the list of user-actions. The clickstream is typically modeled as a directed graph, where web pages constitute the vertices, and edges represent the movement of a user through web pages, i.e., when the user clicks on a hyperlink.
The right-hand side of Figure~\ref{fig:classificationProblem} illustrates the clickstream extracted from the navigation example. Two {\it components} are present: user-actions 2--3--4 are reached following hyperlinks starting from user-action 1, while user-action~5 is a web page reached independently.

We assume a monitoring infrastructure exposes HTTP logs. Examples of such infrastructure are web proxies and network probes that extract information from packets crossing a measurement point. HTTP logs contain records about HTTP requests. Each record includes (i) the time of the request; (ii) the requested URL; (iii) the HTTP \texttt{referer}, i.e., the HTTP header that identifies the URL of the web page that linked to the resource being requested; and (iv) the user agent, i.e., an identification of the browser application sending the HTTP requests.

HTTP does not specify any mechanism for signaling if a request is a user-action or not. As such, HTTP URLs are indistinctly mixed in HTTP logs. Thus, the first problem we target is the identification of user-actions in the stream HTTP requests. URLs identified as user-actions become the clickstream vertices. If a \texttt{referer} is present, it represents a directed edge {\it from} the URL in the \texttt{referer} {\it to} the user-action URL.

As mentioned above, the data we use in our analysis do not provide any user identifier for obvious privacy reasons. We will rely on anonymized IP addresses which, in our case, uniquely identify a \textit{household}, from which several users using different applications and devices may connect. We define the \textit{browser clickstream} as the user-actions performed by a particular {\it browser}\footnote{We use the term browser to refer to any specific application that uses HTTP to fetch resources from the web, e.g., traditional web browsers, mobile apps, etc.} in a household, and thus characterized by the pair of anonymized IP address and user agent string. Obviously, a browser clickstream is not equivalent to the entire activity of a physical person since the same person could use different browsers.

At last, while the user agent string identifies a specific application and version, it also exposes the operating system and type of device being used. We here coarsely group browser clickstreams into three classes of browser devices, namely PCs (including desktops and laptops), tablets, and smartphones.

\section{Related work }
\label{sec:related}

\subsection{Identification of user-actions}

The user-action detection problem we address is similar to web page view identification, a part of the data cleaning process in web usage mining~\cite{Sri00}. Web usage mining is historically the task of extracting knowledge from HTTP server-side log files. As such, this task was traditionally tailored on a per website basis. For instance, web page view identification from web server logs leverages the a priori known structure of the website, and is performed often by discarding a manually constructed list of embedded object extensions from the logs~\cite{S15,SU13}.
This approach does not work in our setting because we aggregate logs from a variety of heterogeneous web servers with different website structures, naming conventions, Ad networks and CDN providers. It is thus not feasible to manually construct a list of extensions to discard.

Previous works have introduced different methods for identifying user-actions from network-based HTTP traces. \mbox{StreamStructure}~\cite{ihm_towards_2011} exploits the \texttt{referer} field in HTTP requests and the widespread deployment of the \emph{Google Analytics beacon} to reconstruct web page structures from HTTP traces and identify user-actions. The authors of~\cite{xie_resurf_2013} follow a similar approach, exploiting the \texttt{referer} field to group requests into HTTP streams. A series of manual rules are used to decide whether a request is the consequence of a user-action or not. Precision and recall above 90\% are claimed on synthetic traces. Finally, the authors of~\cite{houidi_gold_2014} present a heuristic to identify user-actions that operates only with the HTTP requests. The proposed heuristic is shown to scale well in high-speed networks, claiming 66\%--80\% precision and 91\%--97\% recall, depending on parameter choices.

In our previous work~\cite{VD16}, we have introduced a machine learning approach to detect user-actions. In contrast to previous efforts, our approach is fully automatic, does not require manually tuned parameters and is validated with traces from real users.
The present work extends~\cite{VD16} by validating the methodology on  mobile browsers and by using a long-term dataset of HTTP logs collected in an operational network.

\subsection{Web characterization and clickstream analyses}

In the last decade, several works focus on the behavior of end-users -- \eg to determine how often web pages are revisited and how users discover and arrive to web pages. The authors of~\cite{weinreich_not_2008, obendorf_web_2007} characterize web usage exploiting few volunteers' browsing histories. They find that navigation based on search engines and multi-tabbing are changing the way users interact with browsers -- \eg direct web page visits based on bookmarks and backtracking are becoming less popular. A similar study is presented in~\cite{sellen_how_2002}, based on device instrumentation and a small user population, and in~\cite{adar_large_2008}, based on proxy logs.
The closest to our work is~\cite{kumar_characterization_2010}, which leveraged the Yahoo toolbar to analyze web page views from a large population of heterogeneous users over a period of one week. Some of our findings confirm theirs (e.g., deep browsing after leaving search engines) while others present different figures compared to what they discovered (e.g., we observe a higher weight of social networks in referral share and less no-referer traffic). However, with the very fast evolution of the networking technologies and the web, the question whether these results still hold nowadays is raised. Our study answers it on various aspects.

With the emergence of connected mobile devices, the works focusing on mobile users' behavior have multiplied.
The authors of \cite{cui_how_2008} propose a taxonomy of usage of the Internet for mobile users, where data are extracted thanks to contextual inquiries with volunteers. They retrieve three already known categories (\textit{information seeking}, \textit{communication} and \textit{transaction}) and identify a new one: \textit{personal space extension}, anticipating the wide usage of cloud storage systems. The authors of \cite{bohmer_falling_2011} focus on the usage of mobile applications, showing that sessions are generally very short (less than one minute on average).
The authors of \cite{gerpott_empirical_2014} perform a survey on mobile Internet behavior, concluding that the approach to measure the usage (passive/active, objective/subjective, etc.) could heavily impact the results.
A recent work~\cite{ren_analyzing_2017} analyses the Wi-Fi access logs of a city shopping mall, showing that the user revisit periodically the same web content.
Considering HTTPS usage on mobile devices, the authors of~\cite{finamore_mind_2017} study HTTP/HTTPS deployment in mobile networks, finding that HTTPS is mainly used by large internet services. Our work confirm many of these trends. However, we not only present aggregated statistics about protocol usage, but also extract clickstreams from HTTP traffic.

The comparison between mobile and PC Internet usage gave rise to a lot of studies. For example, the authors of~\cite{papapanagiotou_smartphones_2012} present a comparison of objects retrieved from PCs and smartphones, and implications for caching, but without distinguishing  user-actions. The authors of~\cite{tossell_characterizing_2012} compare smartphone and PC navigation, concluding that web page revisits are rare in smartphones, while bookmarks are more widely used on smartphones than on PCs. 
The authors of~\cite{song_exploring_2013} study the differences in searching behavior of mobile, tablet and PC users. They show that most clicked websites depends on the used device, suggesting that these differences should be taken into account to design specific ranking systems.

On a more sociological side, the authors of~\cite{pearce_digital_2013} note that the computers can increase the use of \textit{capital enhancing} activities and to palliate the unavailability of Internet access using mobile devices is not enough to restrain the digital divide.  The authors of~\cite{oulasvirta_habits_2012} note that mobile usage induces a \textit{checking habit} for smartphone, consisting of quick and repetitive content inspection.

Considering clickstream analyses, the work presented in~\cite{huberman_strong_1998} studies the number of web pages that a user visits within a single website, while the one in~\cite{kammenhuber_web_2006} analyses the relations between queries on a search engine and followed paths. On another side, the authors of~\cite{tikhonov_what_2015} define navigation profiles considering data exported by a browser toolbar in Russia, showing that the navigation path leading users to web pages characterizes properties of the destination web page. Finally, the authors of~\cite{olmedilla_mobile_2010} focus primarily on the categories of websites that mobile users access.

In contrast to our work, all these previous works rely on somehow limited vantage points. Here we capture HTTP traces from a large ISP to extract clickstreams, considering different devices and users while connected at residential networks. Moreover, thanks to the three years of data capture, we provide a comprehensive analysis on how browsing behavior is evolving over time. Equally importantly, we study more aspects like the graph structure of clickstreams and the comparison between OSNs and SEs.

The authors of~\cite{xie_resurf_2013} present analyses of clickstreams from passive traces collected for two months. Our work reappraisals some results in~\cite{xie_resurf_2013} -- \eg we confirm that clickstreams are small and restricted to few domains. However, the duration and richness of our data shed light on novel aspects of browsing habits. For instance, in contrast to the previous work, we show that browsing from mobile devices is getting more and more frequent, suggesting that it will soon surpass PC browsing even for users connected at home networks.

The web graph can be obtained thanks to active crawlers~\cite{broder_graph_2000,meusel_graph_2014,bai_discovering_2011}. Our work is orthogonal to those efforts, since they miss how users interact with the web. Contrasting the full web graph with the portion effectively visited by users is planned for future work. Our work can be of interest for the authors of~\cite{wang_unsupervised_2016}, who introduce a system to cluster clickstreams, aiming at mining knowledge from them. Our dataset is a unique source for such analyses, since it covers a large user population for a long period.
\section{User-action classifier design}
\label{sec:validation}

Our classification problem consists in identifying a URL as either user-action or automatic-action. In the past, this problem has been faced by designing ad-hoc heuristics driven by domain knowledge, e.g., by rebuilding the  web page structure~\cite{ihm_towards_2011,xie_resurf_2013}, or manually building blacklists and simple tests~\cite{houidi_gold_2014}. We here revisit the problem and introduce a machine learning methodology. It is given a labeled dataset where the classes of observations are known. Observations are characterized by {\it features}, \ie explanatory variables that describe observations. The classifier uses the knowledge of the class to build a {\it model} that, from features, allows it to separate objects into classes. In the following we summarize our approach that we presented earlier in a workshop~\cite{VD16}.

\subsection{Feature extraction}
\label{sec:proposed}

Instead of a priori selecting features that we believe might be useful for classification, we follow the best practice of machine learning and extract a large number of possibly generic features. We let the classifier build the model and automatically choose the most valuable features for the goal.

Table~\ref{tab.features} lists the features extracted from traffic traces. We consider 17 features that can be coarsely grouped into four non-independent categories: (i)~based on referring relations among URLs; (ii)~based on timestamps; (iii)~describing properties of objects; and (iv)~describing properties of URLs. Features are sorted by their {Information Gain} (IG), a notion that we discuss later.
Some are inspired from prior works.
Some are boolean or categorical, i.e., they can take a limited number of labels, while others are counters or real-valued.

\begin{table}[!tb]
  \centering
  \caption{Features and IG with respect to the {user-actions}.}
  \label{tab.features}
  {
    \small
    \begin{tabular}{|c|c|c|c|c|c|c|}
      \hline
      Feature & \texttt{referer} & Time & Object & URL & Type & IG\cr
      \hline
      Number of Children~\cite{houidi_gold_2014}\cite{xie_resurf_2013}\cite{ihm_towards_2011}         & x &   &   &   &  Count & 0.2706 \cr
      Content Type~\cite{houidi_gold_2014}\cite{xie_resurf_2013}\cite{ihm_towards_2011}        &   &   & x &   & Cat. & 0.0287 \cr
      $\Delta_t$ -- Previous Request &   & x &   &   & Real &0.0140 \cr
      HTTP Status Code~\cite{xie_resurf_2013}              &   &   & x &   & Cat. & 0.0061 \cr
      URL length                     &   &   &   & x & Count &0.0060 \cr
      $\Delta_t$ -- Sibling          & x & x &   &   & Real & 0.0048 \cr
      Ads in URL                     &   &   &   & x & Bool & 0.0040 \cr
      $\Delta_t$ -- Parent~\cite{xie_resurf_2013}\cite{ihm_towards_2011}         & x & x &   &   & Real & 0.0036 \cr
      Content Length~\cite{xie_resurf_2013}                 &   &   & x &   & Count & 0.0027 \cr
      Parent Status Code             & x &   & x &   & Cat. &0.0016 \cr
      Has \texttt{referer}?          & x &   &   &   & Bool &0.0014 \cr
      Max $\Delta_t$ -- Child        & x & x &   &   & Real & 0.0010 \cr
      Parent Content Type            & x &   & x &   & Cat. & 0.0007 \cr
      Ads in \texttt{referer}        & x &   &   & x & Bool & 0.0005 \cr
      Max Length -- Child            & x &   & x  &   & Count & 0.0005 \cr
      Min $\Delta_t$ -- Child        & x & x &   &   & Real &0.0003 \cr
      Parent Content Length          & x &   & x &   & Count& 0.0002 \cr
      \hline
    \end{tabular}
  }
\end{table}

Features are extracted from HTTP logs. Given a URL, we calculate the time interval ($\Delta_t$) from the previous request from the same browser. If the request has a \texttt{referer}, we call the URL in the \texttt{referer} its \emph{parent}. We also determine whether a URL in a request has \emph{children}, i.e., subsequent requests that have this particular URL in the \texttt{referer} field. Based on parent-child relations, we extract the number of children, the time interval between the request and its eventual parent, and the time interval between the request and its last \emph{sibling}, i.e., previous request sharing the same parent. If the request has children, we compute the minimum and maximum time to see a child.

We consider features in server responses, such as the Status Code, Content Type and Content Length. We also augment the feature set with statistics of the request of the parent  (if it exists), e.g., the Content Length, Content Type and Status Code of the parent request. Finally, we include features that describe the URL strings. We count the number of characters in the URL and we check if the URL (or the \texttt{referer}) is included in a blacklist of terms associated with advertisement domains.

\subsection{Classifier choice}

Given the heterogeneity of features and their diverse nature, the choice of which classification model to adopt requires ingenuity. For instance, algorithms based on notion of distance such as Support Vector Machines or nearest neighbor methods are particularly sensitive to the presence of boolean and categorical features. Similarly, the presence of dependencies between the features challenges regression based classifiers. Generally, when there are complex dependencies among features, decision trees and neural networks offer the best performance~\cite{Mitchell_ML}. As such,  we consider:

\noindent {\bf Decision Tree (DT):} It is a tree-like classifier for making sequential decisions on features~\cite{bcart84}. Internal nodes represent tests, branches are the outcomes of tests, and leaves represent classes. We use \texttt{J48} -- an open source implementation of the \texttt{C4.5} decision tree training model.

\noindent{\bf Random Forest (RF)~\cite{Br2001}:} It improves and generalizes decision trees. It constructs a multitude of decision trees at training time using subsets of features, outputting the class that is the mode among those trees. RF is more robust to over-fitting.

\noindent {\bf Multi-Layer Perceptron Neural Network (NN):} It is a feedforward neural network that maps features into classes~\cite{Ha94}. It consists of multiple layers of nodes in a directed graph, where each node is a processing element with a nonlinear activation function. Training is performed with the backpropagation algorithm.

We use the implementations offered by Weka in our experiments.\footnote{http://www.cs.waikato.ac.nz/ml/weka/}

\subsection{Performance metrics and methodology}

The classification performance measures the ability to correctly return the class of an object.
Performance is typically summarized using $Accuracy$, i.e., the fraction of objects from any class that are correctly classified.
Accuracy is often misleading, especially when object classes are unbalanced, i.e., a naive classifier returning always the most popular class would achieve a high accuracy. In such cases, per-class performance metrics must be considered.
Therefore, given we are interested in user-action classification, we also evaluate performance metrics related to this class:
(i) \mbox{$Precision$}: the fraction of requests correctly classified as user-action (the number of true positives among the requests that the classifier selected as user-actions);
(ii) \mbox{$Recall$}: the fraction of user-actions that the classifier captures (number of detected user-actions over the total number of user-actions);
and (iii) $F-Measure$: the harmonic mean of Precision and Recall.

We use the standard stratified 10-fold cross-validation to measure model performance and select the best classification setup, \ie the best tuning parameters.
\section{User-action classifier performance}
\label{sec:clasPerf}
We now provide a performance evaluation of the user-action classifier on ground truth traces. Instead of only building such traces in a controlled environment as done in previous works, we also rely on real traces from actual end-users to build the ground truth.

\subsection{Annotated dataset for training and testing}

Training a classifier and testing its performance require data in which the ground truth is known, i.e., requests are annotated with class labels. In our scenario, we need HTTP logs in which requests are annotated as user- or automatic-actions.

To obtain the ground truth of user-actions, we rely on volunteers. Specifically, we collect the visited web pages from volunteers' PCs by extracting their browsing history from three major browsers: Safari, Chrome and Firefox. Referring to Figure~\ref{fig:classificationProblem}, a volunteer's browsing history exposes the timeline of user-actions. It includes (i)~timestamps of web page visits; (ii)~the requested URLs; and (iii)~codes describing web page transitions -- \eg whether the visit resulted in a redirection to another web page. In total, we observed more than 12\,000 visits to more than 2\,000 websites in 3 months of browsing activity of 10 volunteers.

During the same period, we instrumented our campus network to passively collect the raw HTTP logs of these volunteers by observing the traffic flowing in and out of our university.
We use Tstat~\cite{finamore_experiences_2011}, a passive monitoring tool to perform the collection. The tool exposes information from both the TCP and HTTP headers, including (i) TCP flow-ID, (ii)~timestamps; (iii)~requested URLs; (iv)~user agent; (v)~\texttt{referer}; (vi)~content type; (vii)~content length; and (viii)~status code. Referring to Figure~\ref{fig:classificationProblem}, Tstat exposes the timeline of \emph{all} HTTP requests observed in the network. All in all, we record more than 0.6 million HTTP requests related to the volunteers.

We next match entries in HTTP logs with entries extracted from browsing histories to label user-actions. The matching of entries however requires care. We primarily use the URL and timestamps as keys, but web  page redirections may create issues. For instance, shortened URLs or web  page redirections are logged in different ways by various browsers. We decide to label as user-action the last request in a redirection chain that is present in both HTTP logs and browsing histories. At the end of this process, we mark about 2\% of all HTTP requests as actual user-actions.

\subsection{Feature relevance}

The central idea when doing feature selection is that the data may contain irrelevant or redundant features.
To check which features are relevant to separate user- from automatic- actions, we compute the Information Gain, also known as the mutual information. It quantifies the reduction in entropy caused by partitioning the dataset according to the values of the specific feature. In Table~\ref{tab.features}, we rank features: the higher the information gain, the higher is the information about the user-action class that the specific feature carries in isolation.

We see that the Number of Children is by far the feature with the highest IG. Content Type is well-ranked as well. These results confirm the intuition of prior work~\cite{houidi_gold_2014,xie_resurf_2013,ihm_towards_2011} that \texttt{referer} relations and the analysis of Content Types strongly help in user-actions detection. Next to these features, we find the time interval ($\Delta_t$) between consecutive requests of a single browser, and some other features that are independent of \texttt{referer}, such as HTTP Status Code and Size of URL.

\subsection{Classification performance}

We evaluate the performance of the different classification models. Table~\ref{tab.models} reports the classification accuracy, and F-Measure, precision and recall for the user-action class in the 10-fold cross validation experiments. These results are obtained using all features listed in Table~\ref{tab.features}. Experiments with the best-ranked features result in minor performance variations, which we do not report for brevity (see~\cite{VD16} for more details). The table also lists results for the heuristic proposed by Ben-Houidi \textit{et al}.~\cite{houidi_gold_2014}, which was manually crafted using domain knowledge.

\begin{table} [!t]
 \centering
 \caption{Performance of the classifiers. F-measure, precision and recall for the user-action class.}
 \subfloat[Cross-validation with PC volunteers' training set.] {
    \small
    \begin{tabular}{|x{2.8cm}|x{1.5cm}||x{1.5cm}|x{1.5cm}|x{1.5cm}|x{0cm}}
        \cline{1-5}
        Model & Accuracy  & F-measure & Precision & Recall \nl
        \cline{1-5}
        Decision Tree & 0.996  & 0.906 & 0.905 & 0.907 \nl

        Random Forest  & 0.996 & 0.912 & 0.891 & 0.943 \nl

        Neural Network  & 0.994 &  0.888 & 0.860 & 0.917 \nl
        Manual Heuristic~\cite{houidi_gold_2014} & 0.988 & 0.784 & 0.711 & 0.870 \nl
        \cline{1-5}
    \end{tabular}
    \label{tab.models}
  } \\
  \subfloat[Decision tree tested on smartphone traces for different apps.] {
    \small
    \begin{tabular}{|x{2.8cm}|x{1.5cm}||x{1.5cm}|x{1.5cm}|x{1.5cm}|x{0cm}}
        \cline{1-5}
        Testing Dataset & Accuracy  & F-measure & Precision & Recall \nl
        \cline{1-5}
        Chrome & 0.999 & 0.935 & 0.935 & 0.935 \nl
        UC browser & 0.996 & 0.863 & 0.923 & 0.810 \nl
        Facebook & 0.998 & 0.931 & 0.902 & 0.961 \nl
        Instagram & 0.998 & 0.930 & 0.890  &  0.973 \nl
        \cline{1-5}
    \end{tabular}
    \label{tab.models.test}
  }
\end{table}

We see in Table~\ref{tab.models} that the accuracy is higher than $99\%$ for the three classification models. Recalling that we have about 98\% of automatic-actions, i.e., classes are strongly unbalanced, a naive classifier that always returns ``automatic'' would have about $98\%$ accuracy. As such, we focus on performance for the user-action class. All three machine learning classifiers deliver very good performance. Neural network has the lowest F-Measure among the three. The \texttt{J48} decision tree F-Measure equals to $90.6\%$, while the Random Forest performs marginally better. Interestingly, all machine learning alternatives have much better performance than the manual heuristic (last line in the table) thanks to our larger set of features, and the higher variability in actual user browsing habits that we used to train our models. Recall that the manual heuristic we compare to has been trained and tested on a smaller ground truth trace built in a controlled environment.

Given a decision tree is simpler than a random forest, but with similar performance, we decide to use the former from now on. An appealing characteristic of decision trees is the easy interpretation of the built model. Manually inspecting the tree, we see that indeed features with high IG (\eg number of children) are among the top features of the tree.\footnote{Classifier code can be downloaded from \url{http://bigdata.polito.it/clickstream}.}

\subsection{Training set size}

To get more insights on performance, we run experiments varying the training set. On each round, we consider an increasing number of volunteers in the training set. We then assess performance (i)~using 10-fold cross validation on the same training data; (ii)~validating the model on the remaining volunteers' traces that were not included in the training set, i.e., an independent test set.

\begin{figure}[]
  \centering
  \includegraphics[width=0.43\textwidth]{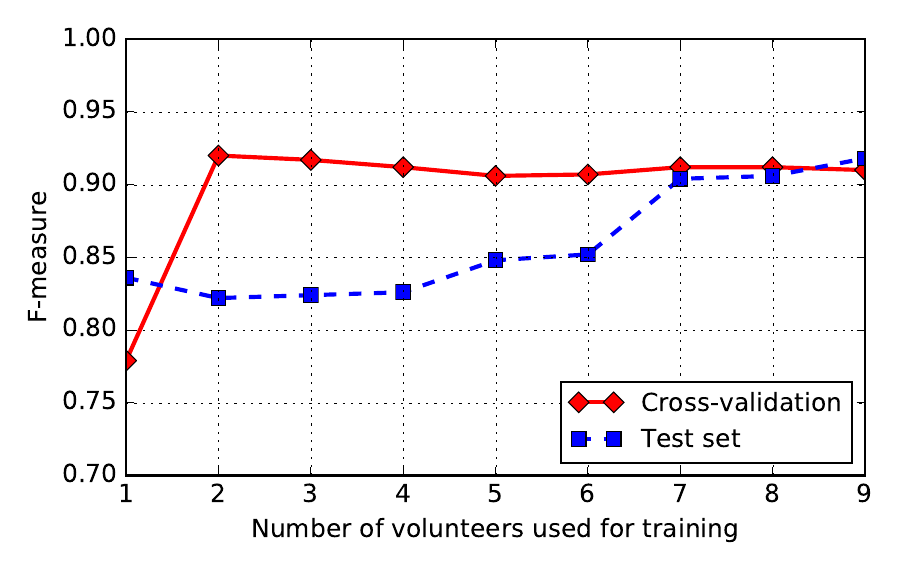}
  \caption{Effects of varying the number of volunteers for training.}
  \label{fig:varying_users}
\end{figure}

The results are on Figure~\ref{fig:varying_users}, where F-Measure for the user-action class is depicted. Considering the cross validation estimates, the F-Measure reaches a plateau when two volunteers are considered. That is, the classifier is able to model the browsing habits of the volunteers included in the training set. More importantly, the validation with independent users shows consistent results, reaching more than 90\% of F-measure when seven or more volunteers are in the training set. In a nutshell, the behavior of the independent volunteers has been learned from other users. These figures provide additional evidence that the produced model is robust and generic.

\subsection{Testing on smartphone traffic}

Training and testing have been done so far considering annotated traces from browsers running on PCs. It is not clear whether the classifier would perform well for smartphone or tablet users too. In particolar we are interested in validating the methodology for mobile browsers, i.e., apps that can be used to navigate through multiple web pages and domains. To answer this question, we collect synthetic traces of visits to popular HTTP websites, taken from top-100 Alexa websites\footnote{https://www.alexa.com/topsites} and by randomly following two links inside each of them.
We performed the experiment by re-opening the URLs with different apps in an Android smartphone, while connected via WiFi to the campus network instrumented with Tstat. We choose four different browsers: Chrome, UC browser, Instagram and Facebook. Instagram and Facebook are apps primarly thought to exploit their internal services, but that allows as well to follow links to external web pages in a in-app browser.
The traces have been collected by visiting these web pages manually, waiting for each web page to be fully-loaded before visiting the next web page. Even if this behavior cannot be considered totally natural as for our PC volunteers, the dataset has some ingredients of real user interactions. Moreover, we do not filter out background traffic of the smartphone.

At the instrumented network, $\approx {5\,000}$-${8\,000}$ HTTP requests have been recorded, depending on the browser, which we classify using the previously described decision tree.
The results are in Table~\ref{tab.models.test}. 
Performances are inline with previous experiments. , \ie precision and F-Measure close to $90\%$. There are very few false positives and even less false negative.  Only UC browser shows a lightly smaller value of recall with respect to the others. Indeed, this browser performs compression of the web pages and some requests are dropped. Therefore, they cannot be recognized as user-actions.

The validation that we performed on different mobile browsers suggests that our machine learning classifier, although trained on PC datasets, adapts well also to traffic towards other devices.

\section{Dataset and traffic characteristics}
\label{sec:datasets}

\subsection{ISP traces}

We evaluate a long-term dataset of HTTP logs captured in a European ISP network to study how users interact with the web and how such interactions evolve over time. Three probes running Tstat have been installed in Points of Presence (PoPs) in different cities, where they observe about 25\,000 households overall. Each household is assigned and uniquely identified by a static IP address. Users connect to the Internet via DSL or FTTH, using a single access gateway offering Ethernet and WiFi home network.\footnote{The sets of households may have slowly changed over 3 years. We however consider always daily and monthly statistics which are marginally affected by such changes.}

Tstat was used to capture HTTP logs, and it saves flow-level statistics simultaneously to the collection of HTTP logs. These statistics include the number of exchanged bytes/packets, flow start/end timestamps, and, for each HTTP request/response pair, timestamp, server hostname, client IP address, URL, \texttt{referer}, user agent string, content type, content length, and status code.\footnote{To reduce the privacy risks, Tstat anonymizes IP addresses and removes parameters from URLs in both \texttt{GET} and in the \texttt{referer} fields.} We rely on the \emph{Universal Device Detection} library\footnote{https://github.com/piwik/device-detector} to parse user agent strings and infer the type of devices (\eg PC, tablet, smartphone, etc.) and the application used. The library operates by matching the user agent strings against a collection of regular expressions describing the different devices.

Tstat also implements DPI mechanisms to identify application layer protocols, such as HTTP and HTTPS. Moreover, Tstat records the server Fully Qualified Domain Name (FQDN) the client resolved via previous DNS queries, using its DN-Hunter plugin~\cite{bermudez_dns_2012}. This mechanism allows us to know which FQDN the client contacted when accessing a given server IP address, and track the usage of HTTP/HTTPS per domain.

We evaluate data collected during 3 years from July 2013 until June 2016. Table~\ref{tab:datasets} summarizes the dataset. In total, Tstat logged information about more than 64 billions of HTTP requests, from which 1.1 billion user-actions are identified. Note that the probes have had some outages during the course of the data collection -- the exact number of days in which each probe was active is shown in Table~\ref{tab:datasets}. We will not show results for the analysis affected by partial outages.

\begin{table}[!tb]
	\centering
	\small
	\caption{Summary of the ISP traces.}
	\label{tab:datasets}
	\begin{tabular}{|x{0.9cm}|x{2.0cm}|x{2.2cm}|x{2.0cm}|x{1.0cm}|x{0cm}}
		\cline{1-5}
		Name  & Households & HTTP Requests & User-actions & Days \nl
		\cline{1-5}
		\PDF & $\approx 10\,000$ & 28.8~billions & 477~millions & 1\,068 \nl 
		\PUL & $\approx 13\,000$ & 30.3~billions & 494~millions &  752 \nl 
		\UMB & $\approx  2\,000$ &  5.3~billions & 79~millions &  600 \nl 
		\cline{1-5}
		Total & $\approx 25\,000$ & 64.4~billions & 1.1~billions & -- \nl
		\cline{1-5}
	\end{tabular}
\end{table}

The dataset captures how users of this ISP interact with the web. Users in different parts of the world will certainly access other domains and services. Thus, some of the results we will present next, such as about top domains promoting content, are certainly specific to this dataset. However, it covers tens of thousands of households and appear representative of the monitored country. For instance, no significant differences are observed among probes.

In the remaining of this section we provide an overview of the dataset and discuss possible limitations, issues and steps we adopt to avoid biases in the analyses.

\subsection{Impact of HTTPS} \label{sec:impact}

Our study is limited to the non-encrypted part of the web. A recent work~\cite{felt2017measuring} reports that HTTPS was responsible for around 45\% of the user-actions by the last month of our data capture. Authors rely on direct instrumentation of Chrome and Firefox. Usage of HTTPS is similar across terminals, but with lower figures on smartphones than on PCs. For example, 38\% (47\%) of the user-actions are over HTTPS for Chrome on Android (Windows) on 4th June 2016. A steady increasing trend on the deployment of HTTPS is observed, but numbers only cover the last year of the ISP traces.

We rely on flow-level statistics saved by Tstat to gauge the effect of HTTPS during the complete duration of the evaluated traces. We quantify how many domains were running over HTTPS and their traffic characteristics (in terms of downloaded bytes). Naturally, we cannot see the exact number of user-actions on HTTPS, but only conjecture how trends reported by~\cite{felt2017measuring} have evolved during the data capture.

\begin{figure}[t!]
    \centering
    \includegraphics[width=0.45\textwidth]{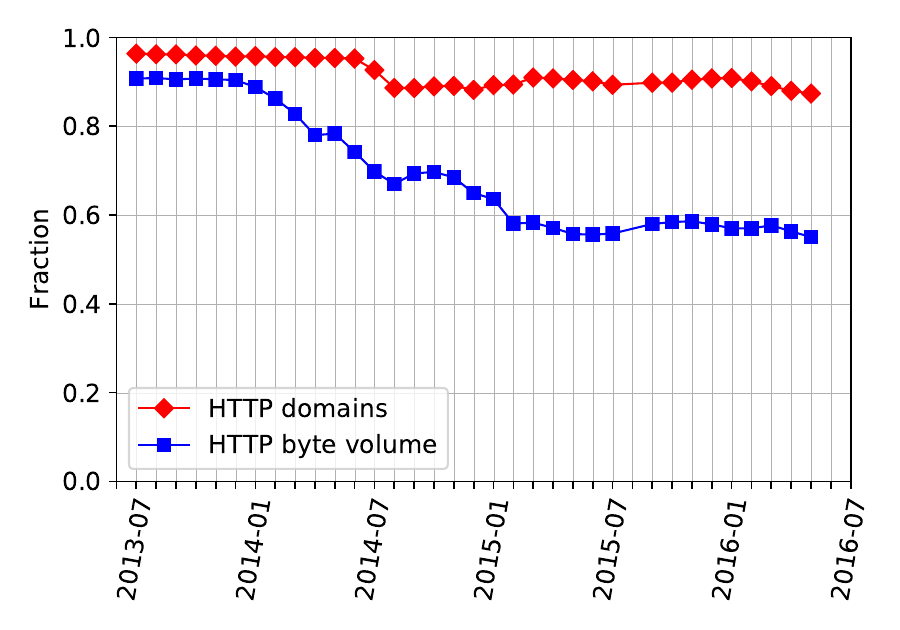}
      \caption{Incidence of HTTP in the dataset. Fraction of domains in HTTP only, and fraction of their byte volumes. Only around 13\% of the domains adopted HTTPS (partly or totally) in June 2016.
    }
    \label{fig:https}
\end{figure}

Figure~\ref{fig:https} shows the share of domains in \PDF relying on HTTP only, \ie completely without encryption. The figure also reports the share of bytes on HTTP. Other probes are omitted since they lead to similar results.

The vast majority of domains relied only on HTTP in 2013 (left-most point in the figure). About $96\%$ of the domains were exclusively running over HTTP, with another $3\%$ using both HTTP and HTTPS. Yet, the overall number of domains using exclusively HTTP remains high -- \eg $87\%$ of the domains, with $6\%$ relying on both protocols in 2016. In terms of downloaded bytes, HTTP was responsible for more than $90\%$ of the traffic in 2013, and the percentage of was reduced to around $55\%$ in 2016 -- \ie $45\%$ of the download traffic was encrypted by the end of our capture.

This discrepancy between domains on HTTP and the HTTP traffic, both in terms of bytes seen in the traces and user-actions reported by~\cite{felt2017measuring}, can be explained by the fact that early-adopters of HTTPS-only deployments are among the most popular domains in the Internet. In fact, observe in the figure the increasing trend on the deployment of HTTPS in 2014, which is related to the migration of YouTube to HTTPS.

\begin{figure*}[!t]
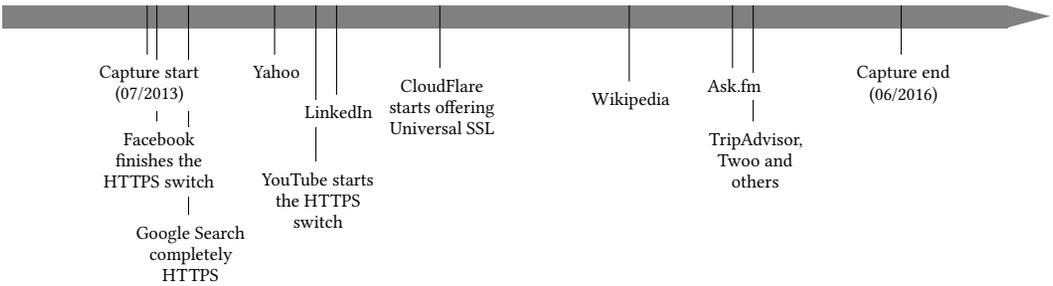

  \scriptsize
	\startchronology[startyear=2013,stopyear=2017,dates=false,color=gray,arrowheight=8px,arrowwidth=15px]
		\chronoevent[date=false,markdepth=70pt,textwidth=1.5cm]{01/09/2013}{Google Search completely HTTPS}
		\chronoevent[date=false,markdepth=10pt,textwidth=1.5cm]{01/01/2014}{Yahoo}
		\chronoevent[date=false,markdepth=50pt,textwidth=1.5cm]{01/03/2014}{YouTube starts the HTTPS switch}
		\chronoevent[date=false,markdepth=35pt,textwidth=1.5cm]{15/07/2013}{Facebook finishes the HTTPS switch}
		\chronoevent[date=false,markdepth=25pt,textwidth=1.5cm]{01/04/2014}{LinkedIn}
		\chronoevent[date=false,markdepth=15pt,textwidth=1.5cm]{01/09/2014}{CloudFlare starts offering Universal SSL}
		\chronoevent[date=false,markdepth=20pt,textwidth=1.5cm]{01/06/2015}{Wikipedia}
		\chronoevent[date=false,markdepth=35pt,textwidth=1.5cm]{01/12/2015}{TripAdvisor, Twoo and others}
		\chronoevent[date=false,markdepth=15pt,textwidth=1.5cm]{01/11/2015}{Ask.fm}
    \chronoevent[date=false,textwidth=1.5cm]{01/07/2013}{Capture start (07/2013)}
		\chronoevent[date=false,textwidth=1.5cm]{01/07/2016}{Capture end  (06/2016)}
	\stopchronology
	\caption{Timeline showing when popular domains running on HTTP in 2013 migrated to HTTPS.}
	\label{fig:timeline_protocols}
\end{figure*}

To further understand which user-actions are missing due to HTTPS, we have studied the domains that were popular in July 2013. For domains presenting a sharp decrease on popularity (more than 60\% reduction on the number of user-actions), we have manually investigated whether they have switched to HTTPS or not. Figure~\ref{fig:timeline_protocols} reports the timeline of  some of the relevant migrations. The traces capture the final periods of Facebook and Google Search migration to HTTPS by default. Other popular domains, such as those from Yahoo and LinkedIn, migrated during 2014. CloudFlare started offering universal HTTPS support for its customers towards the end of 2014. Wikipedia switched to HTTPS in 2015. The migration trend accelerated in the last months of the capture with several (less-popular) domains switching to HTTPS. By the end of the capture, 24 out of the top-100 domains in 2013 have switched to HTTPS.

Further checking the HTTPS traffic, we found that around $85\%$ of the HTTPS bytes in 2016 come from the top-20 domains. Google and Facebook alone account for around $65\%$ of the HTTPS traffic. Relevant for our analysis, a large number of HTTPS domains still pass on the \texttt{referer} information when users transition from the HTTPS domains to any other HTTP domain. This happens because content promoters (\eg Google, Facebook, Yahoo, Twitter etc) have interest in informing others the origin of visits. Technical details are discussed in Appendix~\ref{appendix:a}. Thus, whereas we miss user-actions \emph{inside} HTTPS domains of content promoters, such as Google and Facebook, we still see the information about users' origin when they leave these services. For instance, Google appears as a single vertex in the graph, with edges linking it to web pages visited after users leave its services.

To observe what the migration of services to HTTPS may cause, we set up an experiment, based on trace-driven simulations, to estimate the impact of this. We took all graphs from one full month in 2013 (July), when there was still only a minor portion of HTTPS traffic, and recomputed metrics after removing user actions of an increasing number of websites from the data.~\footnote{Note that pinpointing \emph{all} websites that migrated to HTTPS is not possible. In our simulation, we assume the most popular services are among the first to migrate (which is close to what is observed in reality).} We then incrementally recalculated the metrics until the percentage of missing user actions have reached around 45\% of the total, which is the estimated percentage of missing user actions by the end of our capture in 2016. Notice that this procedure is different from removing 45\% user actions at random, since user actions related to a single site have temporal correlations. Our methodology instead mimics the migration of full popular sites to HTTPS.

The simulation provides more evidences to support the conjectures in the paper. Detailed results will be described in the next sections, alongside the real data.

\subsection{Caveats for the detection of user-actions and clickstreams}

\begin{figure}[!t]
  \begin{center}
    \includegraphics[width=0.44\textwidth]{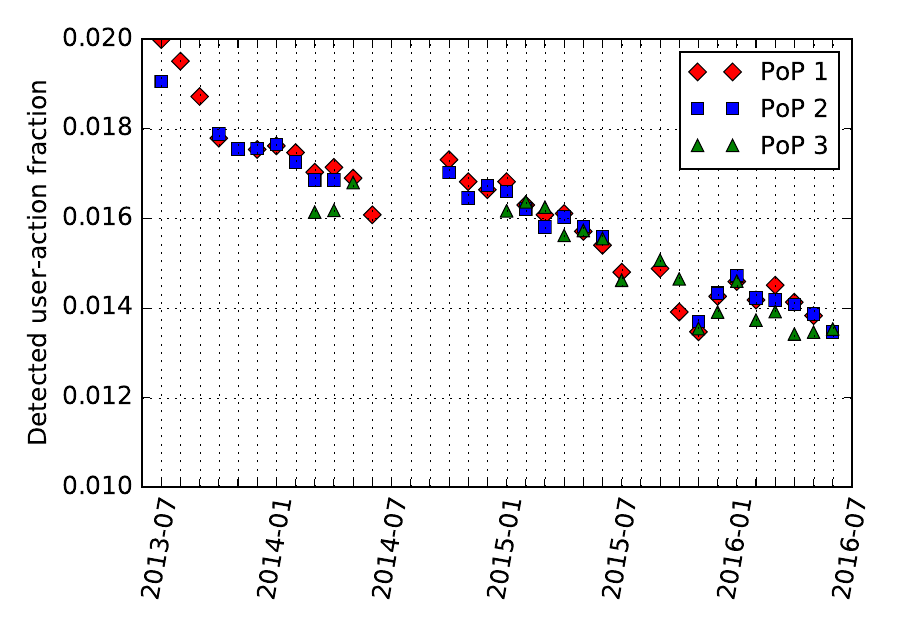}
    \caption{Effects of the increasing complexity of web pages on the percentage of user-actions among HTTP requests.}
    \label{fig:click_percentage}
  \end{center}
\end{figure}

\subsubsection{Effect of web page complexity evolution}

We investigate how the complexity of web pages has changed during the data collection. In particular, given our machine learning approach to detect user-actions, we are interested in checking whether key features have varied significantly throughout the years. Major changes in features would affect the performance of the classification models, which have been trained with data collected from volunteers simultaneously to the last months of the data capture at the ISP network.

We observe that web pages have become more complex in recent years. This observation is inline with previous works~\cite{butkiewicz_characterizing_2014}, which report an increasing trend in the number of objects needed to render web pages. As an example, the median number of children for user-actions is increased by about $40\%$ from 2013 to 2016. As a consequence, the percentage of user-actions among all HTTP requests is decreasing. Figure~\ref{fig:click_percentage} illustrates this effect by depicting the overall percentage of user-actions over time. Observe how similar the trends are across the datasets. The percentage of HTTP requests corresponding to user-actions has decreased from close to $2\%$ in 2013 to less than $1.4\%$ in 2016.

Even if web pages are becoming more complex, the impact on features relevant to our classifier is limited. As an example, the overall mean number of children per HTTP request is more or less constant between 0.71 and 0.75 in three years, and the vast majority of HTTP requests has no children at all. We remind that the number of children is one key feature used by the decision tree to identify user-actions. Similar observation holds for other features used by the trained decision tree.

\subsubsection{\texttt{referer} artifacts}

\begin{figure}[!t]
  \centering
  \includegraphics[width=0.45\textwidth]{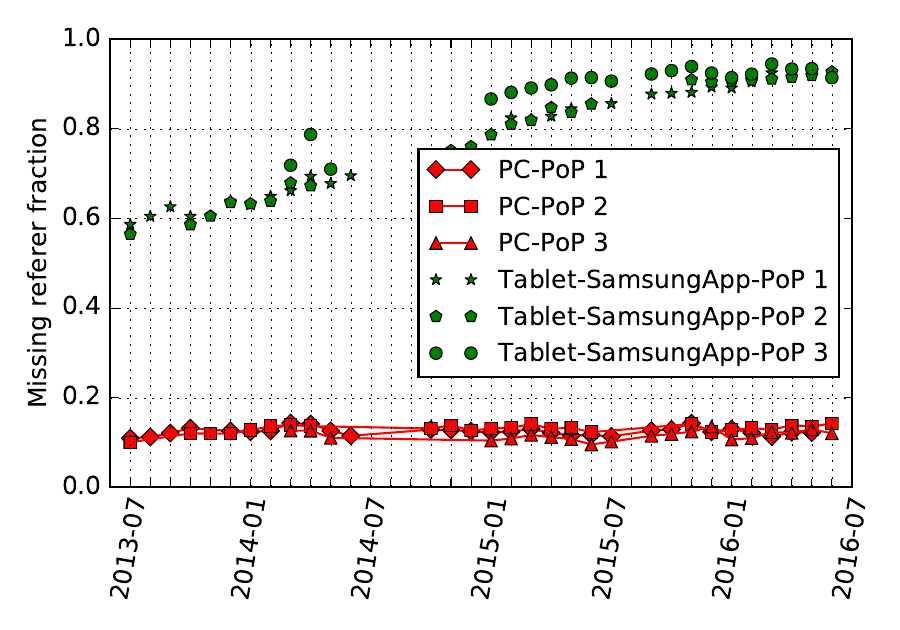}
  \caption{Fraction of missing \texttt{referer} for PC browsers and for the anomalous Android Samsung browser in Tablets.}
  \label{fig:ref_missing}
\end{figure}

The analysis of clickstreams depends on the \texttt{referer} field. Previous work~\cite{schneider_understanding_2009} has reported artifacts related to lack of \texttt{referer} in HTTP requests. Artifacts related to missing \texttt{referer} are expected to be caused by bugged browser implementations or middle-boxes. The latter is not present in our scenario. We then study the number of HTTP requests that miss \texttt{referer} per browser.

Figure~\ref{fig:ref_missing} illustrates the percentages of missing \texttt{referer} when aggregating all browsers running on PCs (red points). Percentages are between $10\%$ and $15\%$ in the three datasets. This behavior is consistent if we check different PC browsers in isolation, as well as most browsers running on smartphones and tablets. This small percentage of missing \texttt{referer} is expected. It is caused by normal browsing activity, such as when users request the first web page after the browser is loaded or when web pages are loaded starting from bookmarks.

However, a completely different picture emerges for few browsers. Figure~\ref{fig:ref_missing} shows that the percentage of missing \texttt{referer} for a specific Android browser running on Samsung Tablets is much higher than in other cases (green points). This behavior is restricted to particular versions of the browser. We see that the percentage of missing \texttt{referer} was close to $60\%$ in 2013 and has continuously grown as more users updated to the versions that skip the \texttt{referer} in HTTP requests. We discard such abnormal browsers in the remaining analyses to avoid biases.

\subsubsection{Clickstreams on mobile terminals}

Some data preparation and filtering is needed to study clickstreams on mobile terminals. Several apps are simply ordinary browsers that behave like PC browsers -- \eg Chrome, Firefox, Safari, Samsung Browser etc. They allow users to move between web pages and domains and pass on the \texttt{referer} information on each transition. Similarly, many apps include their own browsers and allow users to navigate to other web pages and domains without leaving the app. This category includes Facebook, Instagram, Flipboard, Messenger, Gmail among others. Here again, the \texttt{referer} information is passed on normally. Each of these apps sends out a customized user agent string and, as such, they are treated as independent browsers given our definitions (see Table~\ref{tab.nom}).

However, several apps constrain users to few operations, and rely on other browsers (\eg Chrome) or third-party apps (\eg Google Maps) to handle external links. These inter-app transitions are built based on different APIs, and the behavior is not standard across apps and mobile operating systems. As an outcome, the \texttt{referer} information, as observed in the network, when switching between apps is not reliable -- sometimes the \texttt{referer} is an arbitrary string instead of a URL, and often the \texttt{referer} is simply not present.

We have manually evaluated all popular browsers seen in the traces, and ignored browsers that do not allow users to navigate through different pages and domains.

\section{Impact of device on browsing habits}
\label{sec:ContentConsumption}

\begin{figure}[t!]
	\subfloat[PC]{
		\centering
		\includegraphics[width=0.50\textwidth]{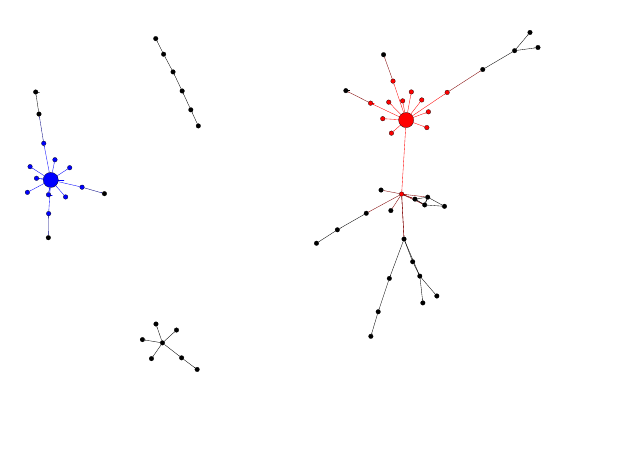}
		\label{fig:left}
	}\hfill
	\centering
	\subfloat[Smartphone]{
		\centering
		\includegraphics[width=0.45\textwidth]{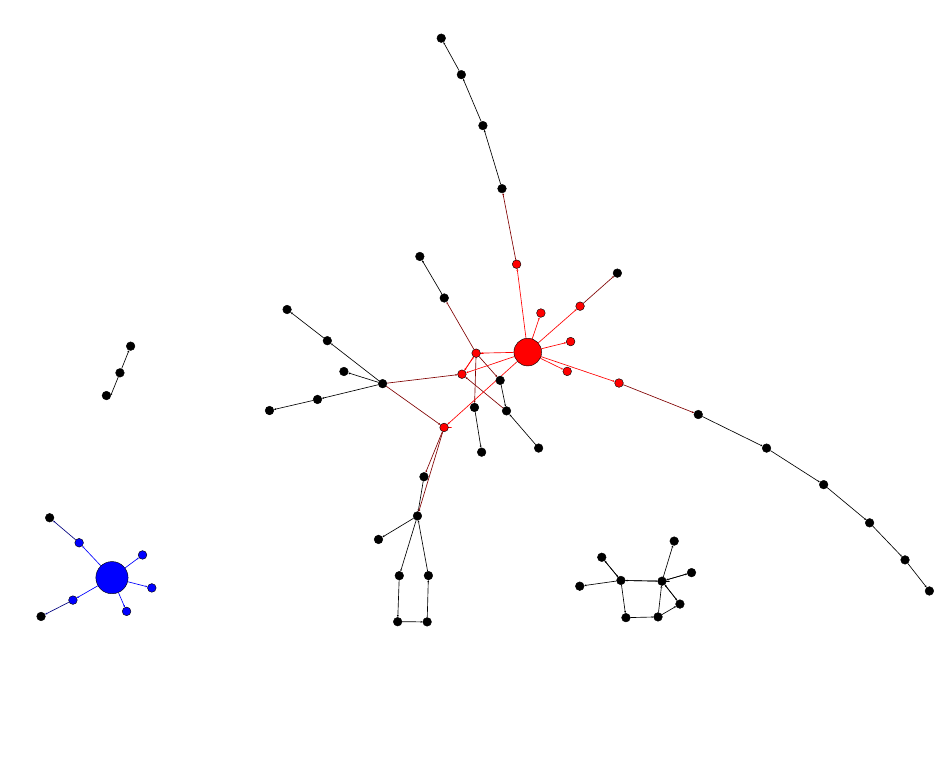}
		\label{fig:right}
	}
	\caption{Browser clickstream samples for a PC and a smartphone browser in the same household in an arbitrary day. SEs and OSNs (and their neighbors) are marked in red and blue, respectively.
	}
	\label{fig:clickstreams}
\end{figure}

We now focus on the analysis of the clickstream graphs to study the long-term evolution of browsing habits. To exemplify the kind of data used in the remaining sections, we show  in Figure~\ref{fig:clickstreams} two cases from an arbitrary household of browser clickstream graphs during one day of navigation. Figure~\ref{fig:left}  refers to a graph in which the user was using a PC to browse the web, while Figure~\ref{fig:right} is related to sessions on a smartphone browser. Bigger colored nodes are SEs and OSNs (red for SEs, blue for OSNs). For instance, Figure~\ref{fig:left} shows four independent components: The bigger starts from web searches from Google; The second one are all pages originated from Facebook; The third and forth components are instead visits not originated by any SE or OSN, and could be possibly due to bookmarks or links received via, e.g., email. These examples already give the intuition about peculiar topologies and the diversity among them. We will quantify such aspects in aggregated and statistical ways in the following  section taking into account all clickstream graphs in the dataset. We will refer again to Figure~\ref{fig:clickstreams} just to illustrate the introduced metrics.

Our aim is to highlight properties of differences in distributions that are of statistical significance. For this, we run a two  Kolmogorov-Smirnov test with the null hypothesis that the two empirical distributions come from the same distribution. The null hypothesis is rejected with a level of significance of 5\%. As a counter-proof, we extract two samples from the same empirical distribution and see if the test does not reject the null hypothesis. 
Given the huge amount of data we are using for our analysis (31 million of user actions per month on average), we observe statistically significant differences even when comparing two empirical distributions that have only small observable differences on the plots. In a nutshell, all statistical tests show significant differences and no statistically significant difference is observed when sampling from the same distribution.

For simplicity, from now on we report results from \PDF only, since the other probes show analogous results.

\subsection{Device usage}

We investigate the evolution in popularity of device categories (PCs, tablets and smartphones). We compute, for each device category, its share in terms of number of active browsers, \ie browsers that generate at least one user-action, and its share in terms of user-actions.

\begin{figure}[!t]
    \centering
    \subfloat[Fraction of active browsers per device category] {
        \label{fig:UA_share_action_a}
        \includegraphics[width=0.45\textwidth]{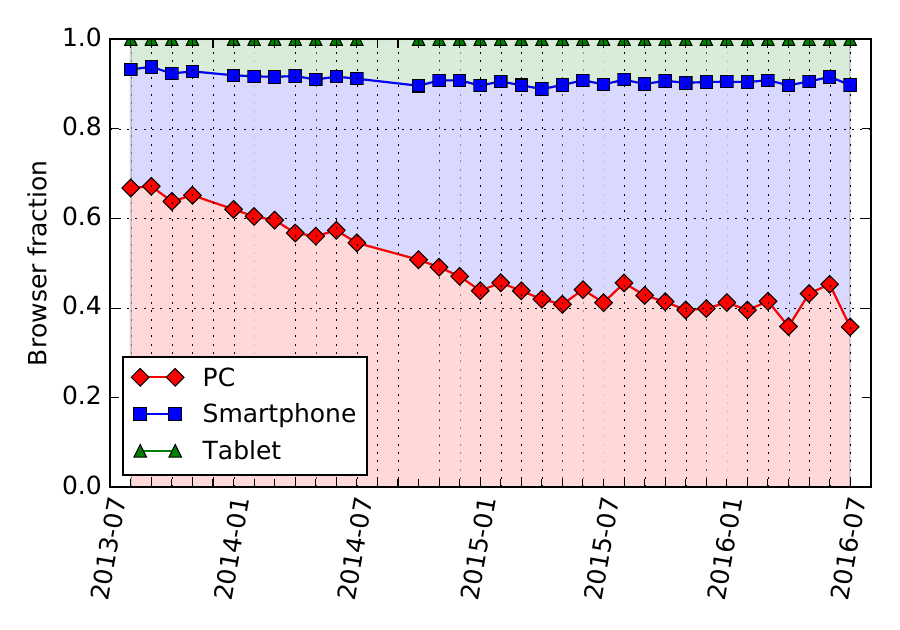}
    }
    \subfloat[Fraction of user-actions per device category] {
        \label{fig:UA_share_action_b}
        \includegraphics[width=0.45\textwidth]{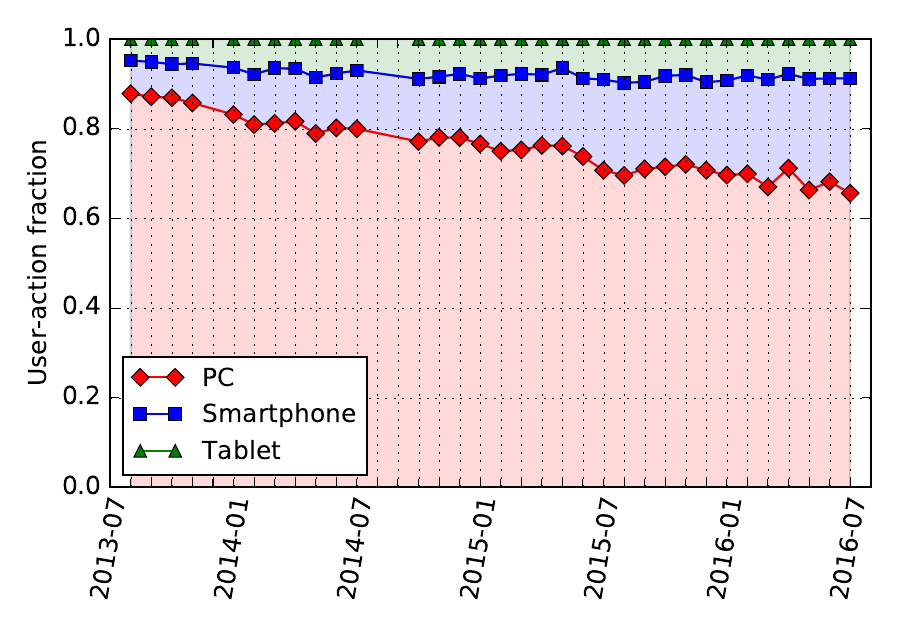}
    }
    \caption{Smartphones present a higher number of browsers, but PCs still dominate the number of user-actions. Browsing from smartphones has increased almost 4 times from 2013.}
    \label{fig:UA_share_action}
\end{figure}

Figure~\ref{fig:UA_share_action_a} shows the fraction of active browsers over the last three years. Observe the stunning increase of smartphone browsers, from $26\%$ to $55\%$ of the total active browsers, with PCs that are now less than $40\%$ of the active browsers.

However, contrast this with Figure~\ref{fig:UA_share_action_b}, which shows the fraction of user-actions per device category.
Although we see an even larger relative growth in the fraction of user-actions coming from smartphones -- \eg from $7\%$ to $27\%$ -- in absolute terms PCs are still creating more user-actions in 2016. For tablets, the increase is more limited, but more visible than in Figure~\ref{fig:UA_share_action_a} -- from  $5\%$ to $9\%$.

Considering the median number of browsers per household, this number has increased from 4 in July 2013 to 7 in June 2016. Smartphones had the largest increase from 1 to 4 (from 1.5 to 6.4 considering the average). PC category remains constant with a median moving between 3 and 4 throughout the years. Tablets are not widespread and the majority of households does not see any browser of this category (with its mean increasing from 0.4 to 1.2). Remember that when an application is updated, we see it as two distinct browsers.

Concluding, we see an increasing number of custom \texttt{user-agent} strings used by different apps in mobile applications, which, recall, we identify as distinct browsers; however, we see a  more limited usage of each browser in mobiles as compared to PCs. We next investigate this latter effect in more details.

\subsection{Browsing sessions}

\subsubsection{Think-time}

\begin{figure}[!t]
    \centering
    \includegraphics[width=0.45\textwidth]{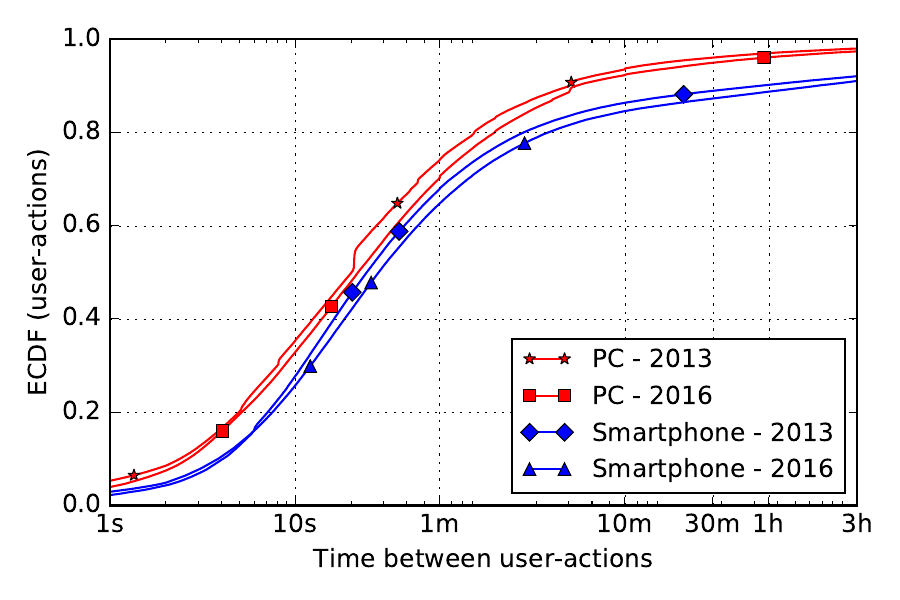}
    \caption{ECDF of think-time for browsers.
     Think-time is shorter on PCs than on smartphones.}
    \label{fig:interarrival}
\end{figure}

Consider the time between two consecutive user-actions, commonly referred as {\it think-time}. Figure~\ref{fig:interarrival} reports Empirical Cumulative Distribution Functions (ECDFs) for smartphones and PCs, comparing July 2013 with June 2016. Tablets are left out to improve visualization.

We observe that in all cases, more than $60\%$ of user-actions are separated by less than 1 minute. The long tail exceeds one day, and peaks are present at typical automatic refreshing time of popular web pages. Think-time is shorter on PCs than on smartphones, suggesting more interactive browsing sessions on PCs.

Interestingly, think-time slightly increased from 2013 to 2016 (median increased of about 10\% for both PCs and smartphones). We hypothesize  that this small increase in think-time could be due to the lack of visibility that the increase in the use of encryption causes to us. In a nutshell, the shift toward  HTTPS reduces our visibility on pages and links which may cause the shift in the think-time we observe. To verify this hypotesis, we  perform the experiment described in Section \ref{sec:impact}. We start from 2013 trace, and remove about 45\% of services as if they migrated to HTTPS. We then recompute the think-time, and obtain a CDF that indeed is very similar to the actual one obtained using 2016 data. We do not report results for the sake of brevity.

\subsubsection{Session time and activity}

We next consider {\it browsing sessions}, \ie the grouped and consecutive generation of user-actions by the same browser. While defining a browsing session is complicated~\cite{Catledge1995,bianco2009web,fomitchev2010google}, we consider a think-time larger than 30~minutes as an indication of the session end. This is a conservative threshold (see Figure~\ref{fig:interarrival}), and it is often seen in previous works (\eg \cite{Catledge1995}), and in applications like Google Analytics.\footnote{\url{https://support.google.com/analytics/answer/2731565?hl=en}}

\begin{figure}[!t]
  \centering
  \subfloat[Browsing session duration] {
      \label{fig:session_a}
      \includegraphics[width=0.45\textwidth]{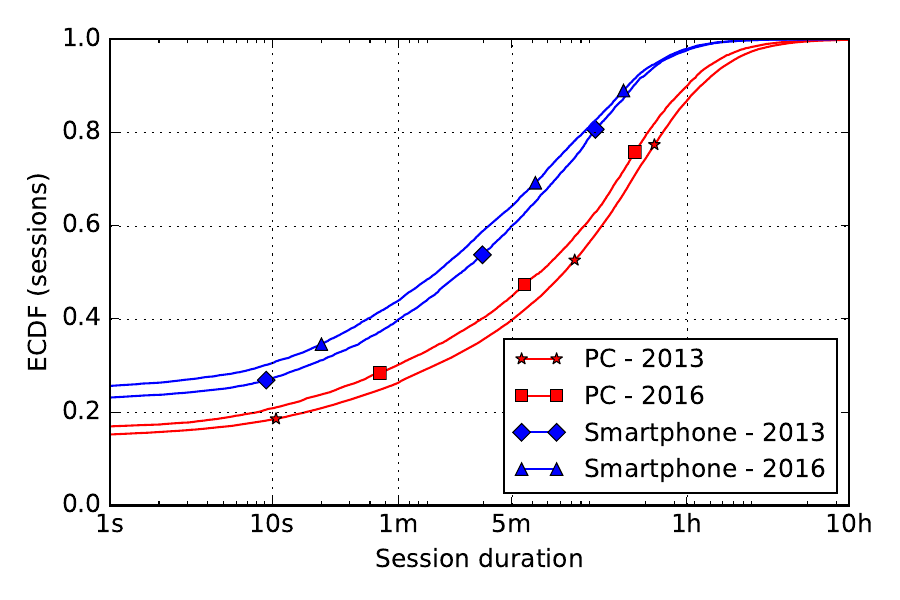}
  }
  \subfloat[Number of user-actions per browsing session] {
      \label{fig:session_b}
      \includegraphics[width=0.45\textwidth]{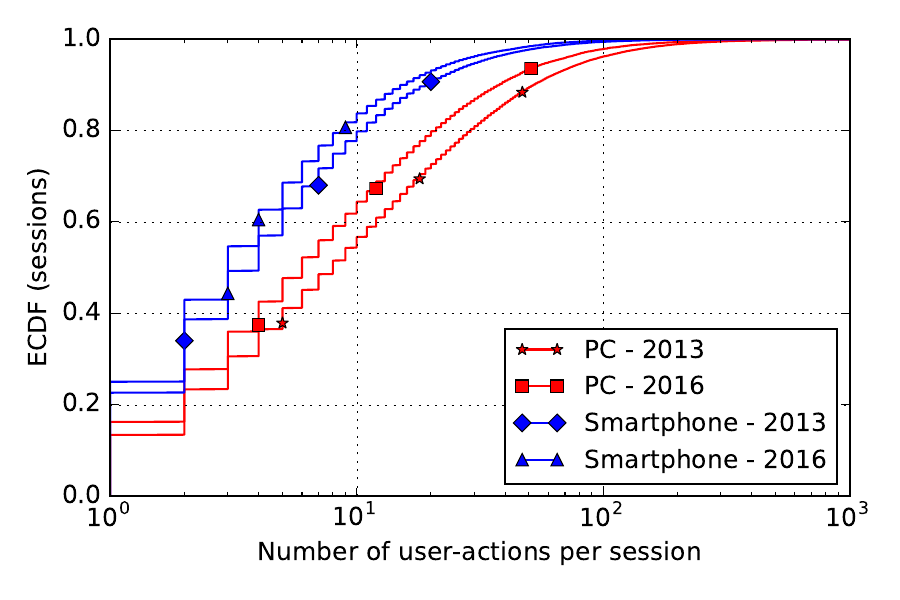}
  }
  \caption{Browsing session characteristics. Sessions on smartphones last shorter with fewer web pages than on PCs. }
  \label{fig:session}
\end{figure}

Figure~\ref{fig:session_a} shows the distribution of session durations and Figure~\ref{fig:session_b} that of user-actions per session. A session with just 1 action is considered of duration 0\,s. Observe that PC sessions last longer and contain more user-actions than smartphone ones.
The median number of user-actions per browsing session is small: half of the smartphone (resp. PC) sessions consist in less than 5 (resp. 9) web pages per session, which do not last more than 2~min (resp. 8~min) in 2016. On the other hand, some few heavy sessions are present: as the tails show, some sessions contain hundreds of web pages and last many hours.

From July 2013 to June 2016 both session duration and number of user-actions per session decreased.
To investigate this change, we run again the experiment of Section \ref{sec:impact}:  from the 2013 trace, we keep removing  websites to the point in which 45\% of the user actions would be missing (due to adoption of HTTPS).
We want to observe if this shift would be compatible with the observed changes in actual data of 2016.

\begin{figure}[!h]
	\centering
	\includegraphics[width=0.45\textwidth]{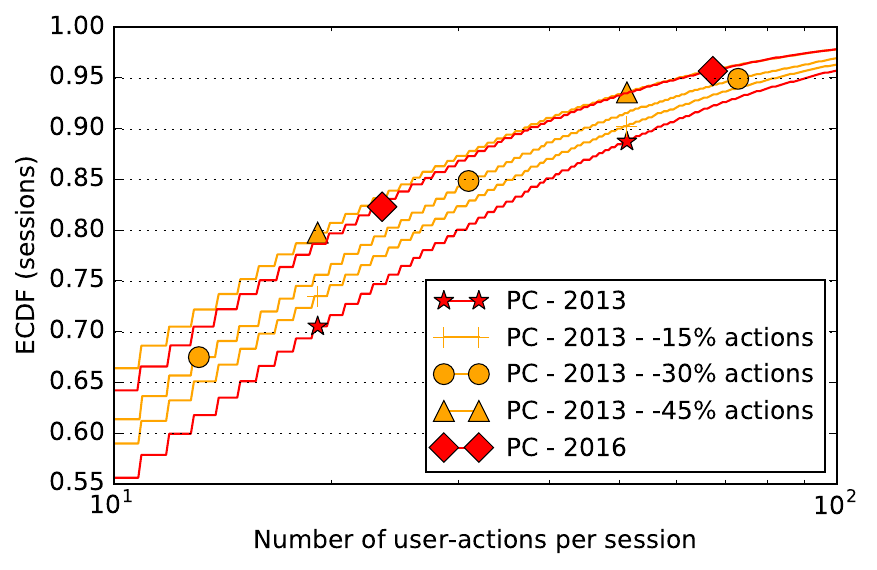}
	\caption{Simulated impact of the HTTPS migration on the number of user-actions per session for increasing percentage of domains that migrated to
		HTTPS.}
	\label{fig:duration2}
\end{figure}

To better highlight this, Figure~\ref{fig:duration2} shows the number of user actions per session, detailing the trend when 15, 30 and 45 \% of services migrated to HTTPS. We consider the PC scenario. As it can be clearly seen, the more services move to HTTPS, the more the curve moves closer and closer to the actual measurements observed in 2016. Notice indeed the tail of the distributions that becomes practically identical.
Whereas this is a not a proof of causality, it provides strong evidences that in this case the differences in three years are likely artefacts due to the shift to HTTPS.

\subsubsection{Inter-session time}

\begin{figure}[t]
    \centering
    \includegraphics[width=0.45\textwidth]{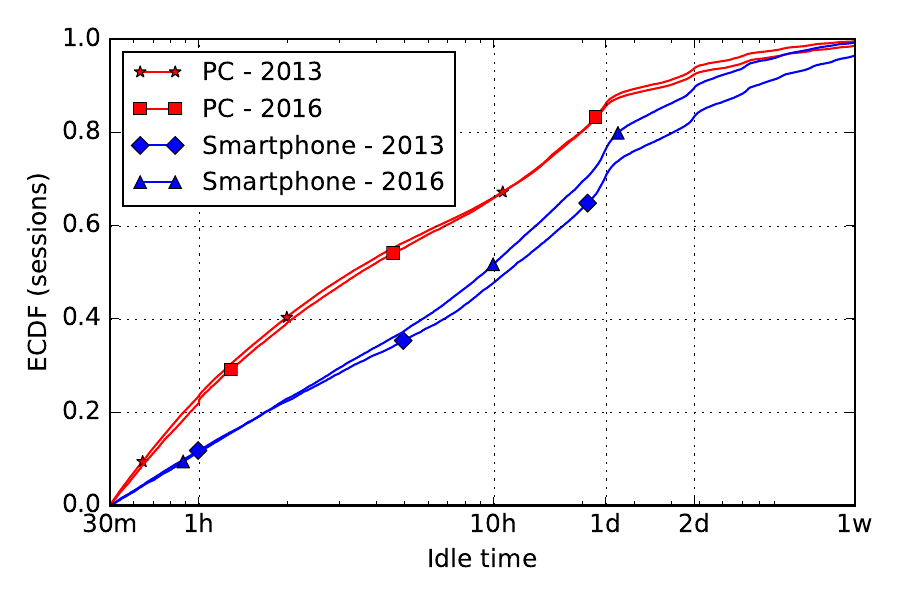}
    \caption{ECDF of session idle time. PC browsers access the web more frequently than smartphone ones, but frequency of use is increasing for the latter.}
    \label{fig:idle}
\end{figure}

To complete the analysis, Figure~\ref{fig:idle} shows the distribution of idle time between sessions. Results are clearly affected by the periodicity of human life: notice jumps at 24 hours, 48 hours, etc. Also in this case, idle time is shorter on PCs than on smartphones, with median values 3 and 7 hours, respectively. We remind that we are considering web page browsing, which is different from other typical usages of mobile terminals, e.g., for instant messaging. Interestingly, smartphone idle time decreased from 2013 to 2016, meaning that the frequency of their usage for browsing the web at home is increasing. The number of times a smartphone browser is used has increased, as well as the number of smartphone browsers.

Again, simulating the migration of services to HTTPS provides further evidences to this claim: HTTPS migration should have caused an increase in the metric. Therefore, the simulation not only reinforces that there has been a decrease in idle time from 2013 to 2016, but also provides an indication about the bias (around 25\%) introduced by the HTTPS migration in the 2016 figures.

In a nutshell, people seem to have short sessions on smartphone and tablet browsers, in which they visit a handful of web pages. Nevertheless, the browsing intensity is increasing on smartphones (as well as the number of used apps), but PCs are still the preferred means for long, concentrated sessions from home.

\subsection{Content consumption}

\begin{figure}[!t]
  \centering
  \includegraphics[width=0.45\textwidth]{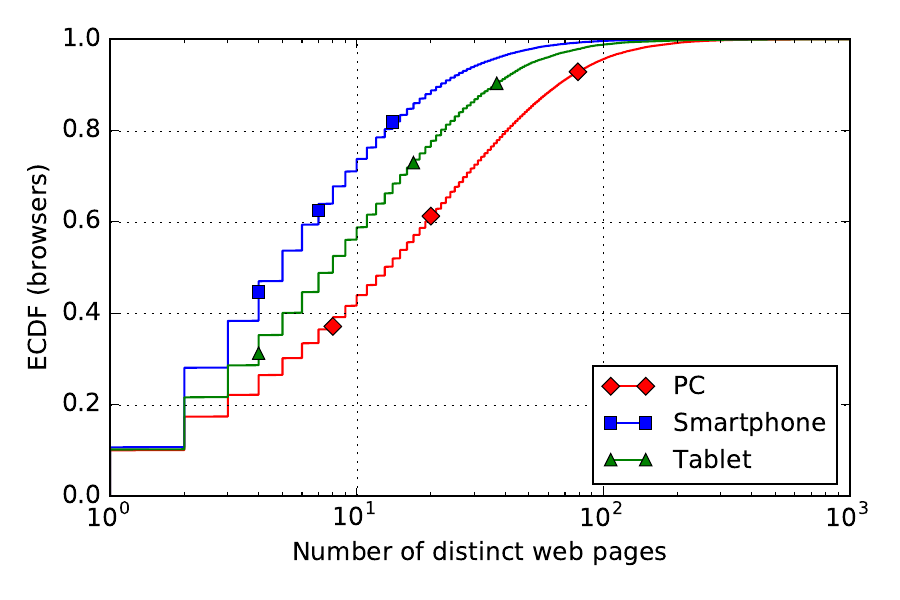}
  \caption{Number of visited web pages per browser per day. PC browsers visit 3 times more web pages per day than smartphone ones.}
  \label{fig:number_nodes}
\end{figure}

Next, we quantify the consumption of content per browser and per device category. Figure~\ref{fig:number_nodes} shows the ECDF of  the number of web pages visited per day per browser, considering all days in June 2016. PC, tablet, and smartphone browsers are depicted in separate lines. As expected, the number of web pages visited on smartphone browsers is significantly lower than the number of web pages visited on PCs, with tablets in between. This reinforces the observation that smartphone browsers are not used for long web page browsing at home.

The total daily number of visited web pages for each browser is in general small: on average, browsers visit 27 distinct web pages per day on PCs,  15 on tablets, and 10 on smartphones. Only few browsers consume more than 300 web pages in total. Comparing the number of user-actions (not shown in the figure) to the number of \emph{unique} visited web pages, we see that each web page is visited $1.5$ times on average.\footnote{Results are overestimated since parameters are removed from URLs.} 
Each browser is seen online only 3.8 days per month (mean values).
Recall that households have a median of $7$ browsers per month (see previous subsections).

\begin{table}[!tb]
	\centering
  \small
	\caption{Details per popular browser application of the average number of visited web pages per day.}
	\label{tab:pop}
	\begin{tabular}{|c||c|c||c|c||c|c|}
		\cline{1-7}
		 \multirow{2}{*}{Browser type}
                  & \multicolumn{2}{c||}{PC}  &  \multicolumn{2}{c||}{Smartphone} &  \multicolumn{2}{c|}{Tablet}  \\
		\cline{2-7}
		  & avg & pop.\% & avg & pop.\% & avg & pop.\%  \\
		\cline{1-7}
		Chrome & 30.9 & 46\% & 9.8 & 39\% & 19.3 & 42\%  \\
		Safari & 23.6 & 6\% & 12.4 & 17\% & 11.8 & 30\%  \\
		Internet Explorer & 16.5 & 20\% & - & - & - & - \\
		Firefox & 30.4 & 20\% & - & - & - & -  \\
		\cline{1-7}
		Others & 22.4 & 8\% & 8.4 & 44\% & 13.8 & 28\%  \\
		\cline{1-7}
	\end{tabular}
\end{table}

To give more details, Table~\ref{tab:pop} reports the average number of visited web pages per day, considering popular browser types for each category (PCs, smartphones and tablets). On PCs, the popular browsers account for 92\% of the observed browsers, with Chrome being the most popular browser type. On smartphones and tablets, instead, there are many more browser types, so that the `others' class accounts for a significant amount of user-actions. Considering user behavior on PCs, we observe that Firefox and Chrome users visit more than 30 web pages per day, while Internet Explorer users are much less active, visiting only 16.5 web pages. On smartphones, Safari users are more active than Chrome users.

\begin{figure}[!t]
	\centering
	\subfloat[PC] {
    \includegraphics[width=0.4\textwidth]{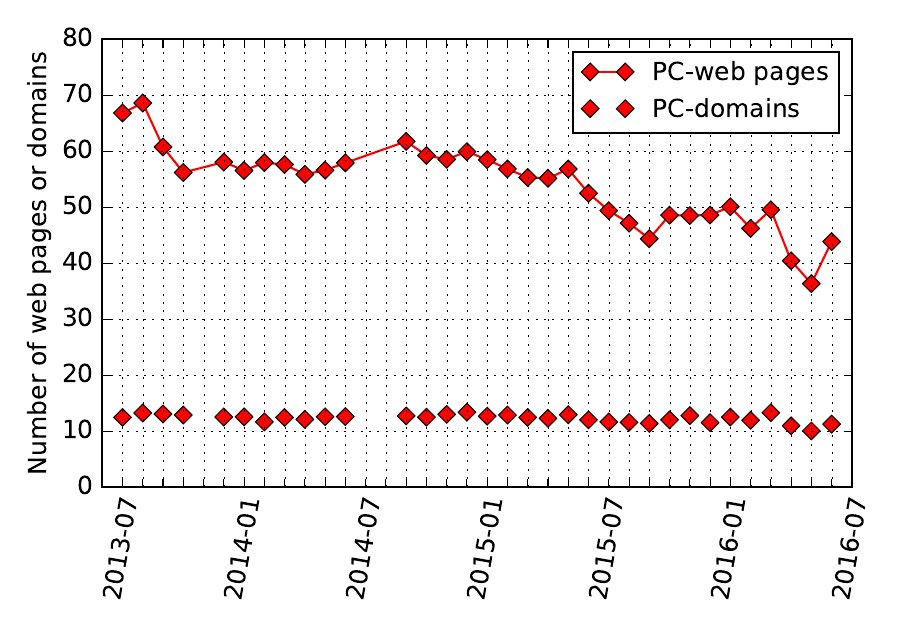}
	}
	\subfloat[Smartphone] {
    \includegraphics[width=0.4\textwidth]{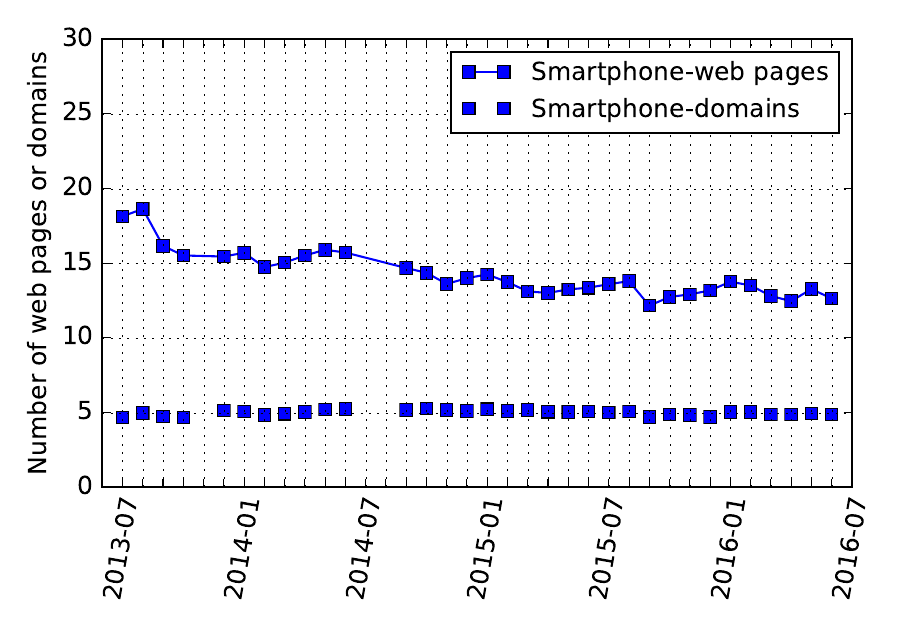}
	}
    \caption{Mean number of web pages and domains visited per day.}
    \label{fig:PDF_pages}
\end{figure}

To determine how content consumption is evolving, Figure~\ref{fig:PDF_pages} shows the evolution of the mean number of daily visited web pages and domains for each browser in the dataset.

Results confirm that Internet browsing habits are in general light: the median number of domains visited in each day is pretty low and constant over time.

We observe a decreasing trend for both smartphones and for PCs. This trend is a consequence of the increasing usage of HTTPS (see~Figure~\ref{fig:https}).
In fact, the figure reports the period in which popular domains started deploying HTTPS (see~Figure~\ref{fig:timeline_protocols}). Google and Facebook had already started their migration before July 2013, but there is still a visible reduction in the series for web pages on PCs, likely connected to the final steps of Google Search migration until September 2013. The migration of some popular domains, such as the Wikipedia, caused a minor, but visible, reduction in numbers for PCs too. The trend is accelerated in the final months of the capture, when other domains starting deploying HTTPS.

To corroborate this claim, we run again the experiment of Section \ref{sec:impact}. If we remove domains that moved to HTTPS in June 2016 from the data of July 2013, the number of daily visited web pages for PCs (smartphones) lowers to 37 (10, respectively). The actual numbers for June 2016 are slightly higher of those numbers, suggesting that the number of web pages visited daily actually did not decrease.

In summary, we conclude that each smartphone browser is used for browsing few web pages at home, while PC browsers are used for more time and to visit more web pages. The number of different domains visited over time is typically small and rather constant.

\section{The clickstream graph}
\label{sec:clickstream}

We now characterize properties of the clickstream graphs looking at how people explore the web over time. For each day, we extract and analyze the clickstream graph for each browser (on average, more than 7\,500 graphs per day), and then compute statistics by aggregating all graphs in each month.

\subsection{Paths characteristics}

First, we gauge how extensive and deep browser explorations are. To reach this goal, we compute how many consecutive and related (by the \texttt{referer} relationship) user-actions, forming a \textit{path}, browsers visit. More precisely, we extracted the longest among the finite directed shortest paths between all vertices in each graph. Such a path is simply called \textit{longest path} further in the paper. Its length gives a hint on how far  the user goes from its navigation starting point over a one-day navigation period.

As seen in the examples of Figure~\ref{fig:clickstreams}, the clickstream graphs are not random graphs, apparently following a preferential attachment structure, with some hubs with a large number of connected web pages and some relatively long branches that form the longest paths we are studying.

\begin{figure}[t!]
  \centering
  \includegraphics[width=0.45\textwidth]{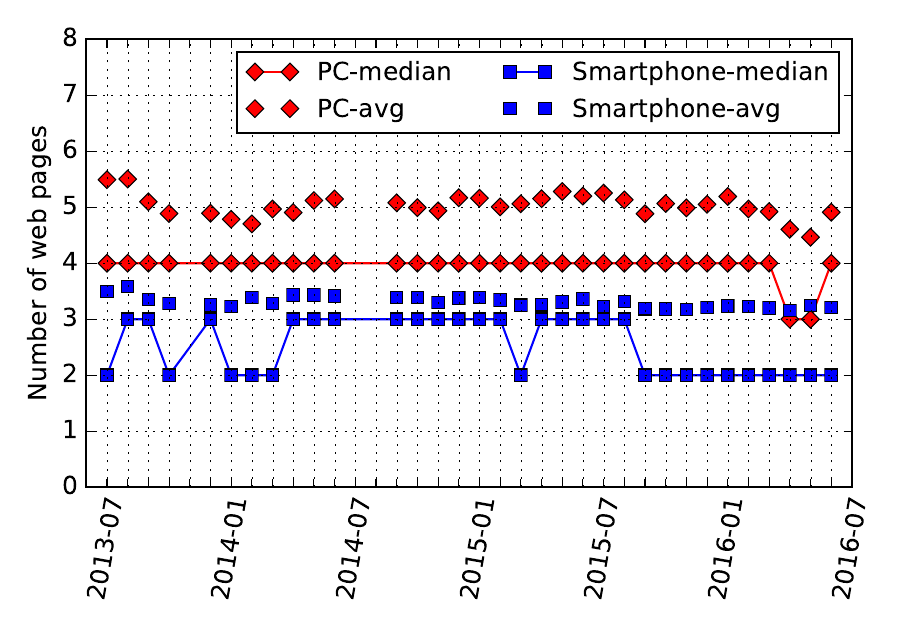}
  \caption{Daily number of visited web pages in the longest path of a clickstream. Depth of the graphs is very limited.}
  \label{fig:page_path}
\end{figure}

Figure~\ref{fig:page_path} shows the evolution of median and average number of user-actions in the longest paths for PC and smartphones. Paths are quite short, and longer on PCs than on smartphones. Interestingly, path length is stable throughout the years, even if the number of web pages is reduced (see PCs in Figure~\ref{fig:number_nodes}).
The migration to HTTPS has little impact here: the longest paths do not appear to be exclusively through encrypted domains, even if their popularity (in terms of user-actions) is quite high.

\begin{figure}[t!]
  \centering
  \includegraphics[width=0.45\textwidth]{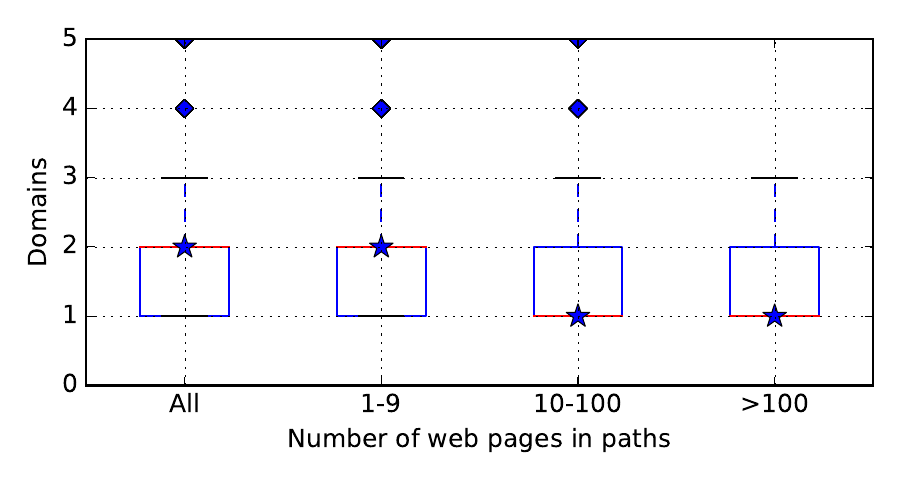}
  \caption{Box-plots of the number of domains in the paths, per path length. Users stay in very few domains, even when navigating through hundreds of web pages.}
  \label{fig:domain_per_url}
\end{figure}

If we count the number of unique domains in each path, we obtain that, on average, only $1.8$ domains are present in the longest path. This suggests that people tend to perform deep navigation in web pages from the same domain, rather than moving among multiple consecutive domains.
One could expect this to be a consequence of the fact that most paths are very short. To check whether long paths differ from short paths, we extract path characteristics conditioned to the number of user-actions in the path, i.e., for paths with less than 10 user-actions, from 10 to 100 user-actions, or with more than 100 user-actions. Figure~\ref{fig:domain_per_url} summarizes with whiskers box-plots the distributions of the number of domains in the paths and it is cut at 6 domains for ease of visualization. Outliers are recognized with the classical Tukey rule on quartile (1.5 IQR method) and shown with diamond markers.  In our case, only 3.8\% of all the paths are outliers, with the remaining 96.2\% of samples that have a number of domains per path that is not greater than 3, with the median number (starred mark) of domains that is always 1 or 2.
Therefore the number of domains in paths is always extremely limited, and this number does not increase for longer paths. Unexpectedly, all paths with more than 100 visited web pages are within 3 domains too. We investigated this behavior and found that long paths are related to peculiar domains, such as comics or galleries with hundreds of images.

In conclusion, user paths among web pages are rather short, and rarely users move through many consecutive domains when navigating.

\subsection{Connected components}

We now study the structure of the overall clickstream graph to check whether the several paths forming a clickstream are independent or connected. To this end, we consider the size of the biggest \emph{Weakly Connected Components} (WCC). A WCC is any maximum sub-graph such that there exists a path between any pair of vertices, considering {\it undirected} edges. In a large WCC, the vertices are connected by edges, i.e., web pages are visited by following hyperlinks. Recalling the examples of Figure~\ref{fig:clickstreams}, in those cases we have 4 and 3 WCCs, respectively for the PC and for the smartphone browser, with the biggest WCCs accounting for the majority of the visited web pages.

\begin{figure}[t!]
  \centering
  \includegraphics[width=0.45\textwidth]{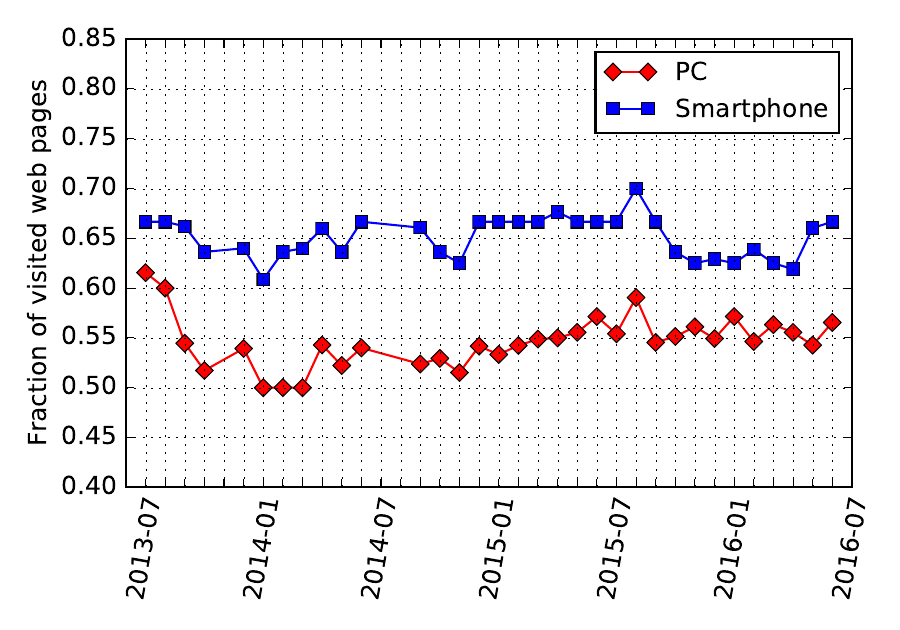}
  \caption{Size of the WCC over time (median over browsers). WCCs typically include SE and OSN. On smartphones, the WCC is bigger than on PCs. This is likely due to the higher usage of SEs and OSNs to reach content.}

  \label{fig:wcc}
\end{figure}

As usual, we report the evolution over time of the measurements. Figure~\ref{fig:wcc} shows the median of the ratio between the {biggest WCC size} and the {entire graph size}, i.e., the relative extension of the WCC.\footnote{Our WCC is a lower bound estimation of the actual WCC: visits to web pages served over HTTPS are invisible to us; the user-action classifier could miss some user-actions, and some \texttt{referer} fields may be missing in requests. These artifacts correspond to missing edges, thus shrinking the largest WCC.} Interestingly, minor changes are observed over the three years, showing minor modification in browsing habits.

Observe that the biggest WCC covers more than $50\%$ of the entire graph, surpassing $70\%$ for smartphones. By manually inspecting the biggest WCCs, we notice that they usually include SEs and/or OSNs, which act as hubs connecting many user-actions. Recalling that paths are usually short this suggests that SEs and OSNs act as starting web pages used by people to reach other content. We will better investigate this in the next section.

We now focus on those vertices which are not connected to the biggest WCC, but form other (small) WCCs. These typically do not contain SEs or OSNs.
A new WCC  can indeed be created when there is no \texttt{referer} in a user-action, i.e., when the user directly visits a web page from a bookmark, or by directly entering the address in the browser bar.
These would be web pages users are familiar with, and they reach without following hyperlinks.
Notice how the number of vertices that are not part of the biggest WCC is higher for PCs than for smartphones. We conjecture that this can be explained by the different usability of devices: PC browsers facilitate the direct access to content (via both bookmarks, and by auto-completing URLs manually entered). On smartphones, users tend to get support of SEs and hyperlinks to reach the desired content, which results in a more extended WCC.

\section{Content Discovery}\label{sec:content_discovery}

In this section we investigate in more details how users reach content on the web, and which are the domains that act as content promoters. We use the out-degree of a domain\footnote{The out-degree of a domain is the number of hyperlinks leaving web pages from this domain to web pages of another one.} as a metric to pinpoint the promoters: the higher this number is, the more different domains are reached from it. We consider all clickstream graphs in June of each year, and, for each domain, we count the total number of unique domains it appears as a parent at least once.

Table~\ref{tab:prom} shows the evolution of the top promoters in terms of the fraction of different domains they promoted. The great majority are SEs and OSNs. Both facilitate the user to look for content, and are the typical way to start browsing activities and to discover new domains.  Google has a dominant position, with more than 6 times more domains promoted than Facebook which comes second. Its leading position has been constant over years even if in slight decrease, with 55\% of domains that have been reached at least one time from \url{google.it} domains in June 2016. We observe little changes in the ranking, with Facebook, Bing and Twitter becoming more popular means to reach content.

\begin{table}[!tb]
	\centering
	\small
	\caption{Ranking of content promoters popularity with the percentage of domains visited from them.}
	\label{tab:prom}
	\begin{tabular}{|l|c|c|c|c|}
		 \hline
Domain & 2013  & 2014 & 2015 & 2016  \\

 \hline
google.it	&	1 (58.3\%) &	1 (56.1\%)&	1 (54.5\%)&		1 (54.5\%) \\
facebook.com	&	3 (6.2\%)&		3 (6.9\%)&	2 (8.1\%)&		2 (9.1\%)\\
bing.com &	5 (2.0\%)&	5  (2.3\%) &		4  (2.8\%) &		3 (4.0\%) \\
yahoo.com &	4  (3.1\%)&	4 (2.6\%)	&	5 (2.4\%) &	4  (2.0\%)\\
google.com  	&2 (6.5\%)&		2  (6.9\%)&		$3$  (3.0\%)	&	$5$  (1.7\%) \\
twitter.com	&	13  ($0.5$\%)&		9 ($0.5$\%) &		8  ($0.6$\%)&		8  (0.6\%)\\
 \hline
  \end{tabular}
\end{table}

\begin{figure}[]
	\centering
	\includegraphics[width=0.46\textwidth]{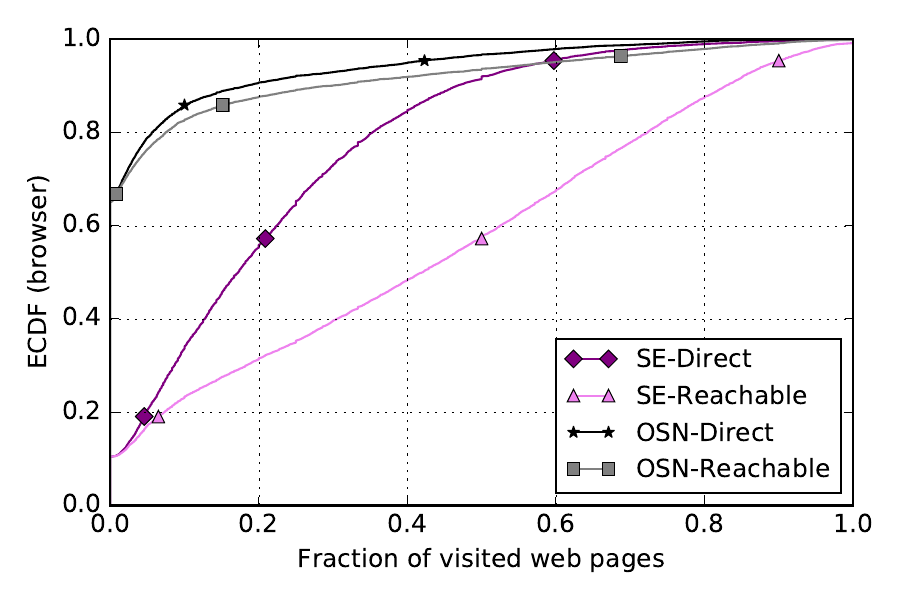}
	\caption{Fraction of distinct web pages per clickstream graph that are directly connected to or reachable from SEs and OSNs. Paths starting from SEs include a sequence of multiple web page visits, while OSNs typically promote a single web page.}
	\label{fig:reach_search_engines}
\end{figure}

\begin{figure}[]
	\centering
	\subfloat[PC]{
		\label{fig:PC_OSN_a}
		\centering
		\includegraphics[width=0.45\textwidth]{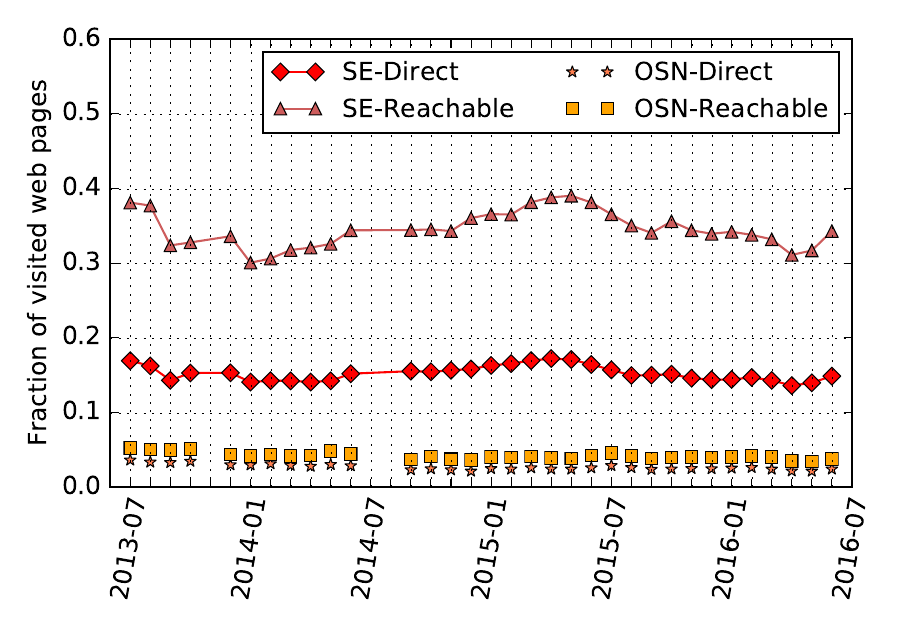}
	}
	\subfloat[Smartphone]{
		\label{fig:PC_OSN_b}
		\centering
		\includegraphics[width=0.45\textwidth]{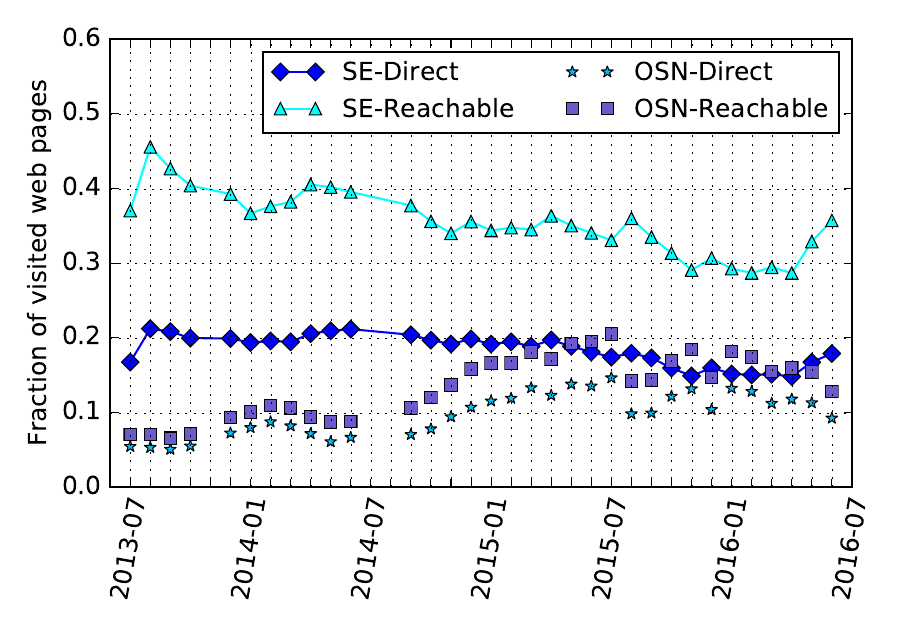}
	}
	\caption{Average number of web pages directly connected to and reachable from SEs and OSNs.
		SEs promote more content than OSNs. On smartphones, OSNs have become as important as SEs to directly reach content.}
	\label{fig:PC_OSN}
\end{figure}

We now focus on individual clickstream graph to better understand (i) how the usage of SEs and OSNs varies across users and devices; and (ii) if users keep visiting other web pages after their first visit from a promoter. We manually build a list of the top-50 SEs and OSNs that promote content. To verify to what extent users explore the web after leaving any of these domains, we define the concept of \textit{reachability}: we say that a destination web page is \textit{reachable} if there exists a direct path in the clickstream graph from the promoter web page to the destination web page. A web page is {\it directly connected} if it is a child of the promoter. We consider June 2016, and browsers that visited at least 20 web pages.

Given a browser clickstream graph, consider the fraction of visited web pages  that are directly connected to (or reachable from) an SE or an OSN.
Figure~\ref{fig:reach_search_engines} shows the ECDFs of this fraction across all graphs.
For instance, consider the SE curves. 10\% of graphs have a fraction of 0\%, meaning that 10\% of browsers have no web page connected to any SE.
The median fraction of web pages per graph which are directly connected to a SE is approximately 17\%, i.e., 17\% of web pages in a browser are found thanks to the direct support of a SE.

Compare now the fraction of web pages reachable and directly connected to SEs. Figure~\ref{fig:reach_search_engines} shows that there are many more web pages that are reachable from search engines (median of $42\%$) than web pages directly connected to them. This indicates that users keep browsing the web after leaving a SE, and keep visiting web pages after the initial one.
OSNs, on the other hand, exhibit a very different pattern. First, the fraction of web pages to which they generate visits is much lower, with about $65\%$ of graphs that contains no web pages visited from any OSNs. Second, the \textit{Direct} and \textit{Reachable} curves are very close to each other, suggesting that OSN users visit some external web page (direct visit), but then do not continue visiting any other web page (indirect visit). Overall, $61\%$ of the web pages connected to an OSN have no other child.
This shows some ``addictiveness'' of OSN users: they click on external links, but then get back to the OSN web page, without continuing exploring the web.

At last, we track the evolution over time of the average direct and reachable fraction of web pages from SEs and OSNs. Figs.~\ref{fig:PC_OSN_a} and \ref{fig:PC_OSN_b} report results for PCs and for smartphones, respectively. PC browsers  show little changes, with SEs being their preferred means to start their browsing paths and to discover content. OSNs are less often used to discover web pages, and very rarely to reach web pages not directly promoted on the OSN itself (cf. Figure~\ref{fig:reach_search_engines}).
Instead, results for smartphones show a clear evolution over time: OSNs are much more important to discover content than for PC browsers. Even more interestingly, we observe a decreasing usage of SEs to look for content. Notice indeed that the fraction of web pages reached from OSNs is now comparable to the one reached from SEs.

\section{Conclusions}\label{sec:future}

Clickstreams offer invaluable information to understand how people explore the web. This paper provides a thorough and longitudinal characterization of clickstreams from passive measurements.

Key challenges for such a study are the identification of URLs effectively requested by users in HTTP traces, and the availability of data summarizing users' activity. We proposed a machine learning approach to identify user-actions in HTTP traces, which we showed to perform well in different scenarios. We then applied it to a large dataset describing three years of activity in thousands of households. We offer the anonymized data to the community in the hope to foster further studies on browsing behaviors. To the best of our knowledge, this is one of the largest datasets available to the study of clickstreams.

Our characterization answers two research questions. First, we quantified several aspects of browsing behaviors and how clickstreams have evolved for 3 years. Second, we uncovered properties of clickstream graphs according to the different types of client devices. We observed interesting trends, such as the typical short paths followed by people while navigating the web and the fast increasing trend in browsing from mobile devices. We observed the different roles of search engines and social networks in promoting content in PCs and smartphones. Finally, we also highlighted how the deployment of HTTPS impacts the study of clickstreams from network traffic. Our results, while sometimes confirming intuitions, precisely quantify the various aspects of clickstreams, with implications for targeted on-line advertisement, content recommendation, the study of on-line privacy, among others.

We envision some promising directions for future work. First, whereas HTTPS prevents the full reconstruction of clickstreams, contacted domains are still visible in the network. We plan to investigate whether machine learning approaches are able to reconstruct clickstreams -- at a domain level -- from HTTPS traffic. Such an approach would provide a coarse view of paths followed by users while surfing the HTTPS web. Second, this study has focused in structural properties of clickstreams without considering the type of content people visit. Characterizing clickstreams according to properties of the visited web pages is a natural continuation.


\bibliographystyle{abbrv}

\bibliography{journal}

\appendix
\section{Encrypted domains and the \texttt{referer}}
\label{appendix:a}

Visits to web pages in encrypted domains are not visible through passive measurements of the network. Moreover, RFC~7231 mandates that ``a user agent MUST NOT send a Referer header field in an unsecured HTTP request if the referring page was received with a secure protocol''. Therefore, \texttt{referer} fields on HTTPS to HTTP transitions should not be visible in the network.

However, RFC~7231 is not strictly followed by popular domains that promote content. Content promoters have incentives to pass on the \texttt{referer} to domains they promote, including non-HTTPS destinations. For illustration, we quote Facebook when it switches to HTTPS in 2013:\footnote{\url{https://www.facebook.com/notes/facebook-engineering/secure-browsing-by-defa\%20ult/10151590414803920}}
\begin{quote}
Browsers traditionally omit the referrer when navigating from https to http. When you click on an external link, we'd like the destination site to know you came from Facebook while not learning your user id or other sensitive information. On most browsers, we redirect through an http page to accomplish this, but on Google Chrome we can skip this redirect by using their meta referrer feature.
\end{quote}

Thus, many different strategies have been adopted over the years when moving from HTTPS to HTTP. Facebook statement explains that modern web browsers include a mechanism to allow HTTPS domains to declare whether the \texttt{referer} field should be passed on or not (the HTML~5 \texttt{meta referrer} tag). By the time of writing, large content promoters rely on the mechanism to pass on some \texttt{referer} information, such as Facebook, Twitter, Bing and Yahoo.

The HTML~5 \texttt{meta referrer} tag is however rather recent. Before it, each content promoter used to implement its own solution to pass on the \texttt{referer} to non-HTTPS domains. In Facebook statement, for instance, we see that for browsers not supporting the \texttt{meta referrer} tag Facebook first redirects users to a Facebook domain that is still running on HTTP. In this first redirection, the original \texttt{referer} is lost as specified by RFC~7231, but users are still in Facebook systems. Then, users are again redirected to the final destination domain. The final destination receives as \texttt{referer} the HTTP domain under control of Facebook.

As we have seen, the majority of domains was not under HTTPS during the data capture. Popular content promoters were the exception, since they were the early-adopters of full HTTPS support. Since content promoters implement the above techniques to pass on the \texttt{referer}, transitions from the content promoters to HTTP domains are indeed considered. Instead, HTTPS to HTTPS transitions are not visible -- \eg internal navigation on HTTPS domains (\eg Google$\rightarrow$Google) or transition between two HTTPS domains (\eg Google$\rightarrow$Facebook). Clearly, as more and more domains move to HTTPS-only, more transitions become invisible.

\end{document}